\documentclass[english]{IEEEtran}
\usepackage[T1]{fontenc}
\usepackage[utf8]{inputenc}
\usepackage[active]{srcltx}
\usepackage{textcomp}
\usepackage{amsmath}
\usepackage{amssymb}
\usepackage{graphicx}
\usepackage{textgreek}

\makeatletter

\newcommand{\lyxmathsym}[1]{\ifmmode\begingroup\def\b@ld{bold}
  \text{\ifx\math@version\b@ld\bfseries\fi#1}\endgroup\else#1\fi}

\providecommand{\tabularnewline}{\\}

\usepackage{cite}
\usepackage{amsmath,amssymb,amsfonts}
\usepackage{algorithmic}
\usepackage{xcolor}
\usepackage{cuted}
\usepackage{flushend} 
\usepackage{url}
\usepackage{tikz}        
\usepackage{circuitikz}  
\usetikzlibrary{arrows.meta} 
\def\BibTeX{{\rm B\kern-.05em{\sc i\kern-.025em b}\kern-.08em
    T\kern-.1667em\lower.7ex\hbox{E}\kern-.125emX}}

\makeatother

\usepackage{babel}
\begin{document}
\author{Moshe Azoulay,
        Gilad Orr,
        and Gady Golan%
\thanks{M. Azoulay and G. Golan are with the Department of Electrical Engineering, Ariel University, Ariel 40700, Israel (e-mail: mosheaz@ariel.ac.il; gadygo@ariel.ac.il).}%
\thanks{G. Orr is with the Department of Physics, Ariel University, Ariel 40700, Israel (e-mail: gilad.orr@ariel.ac.il; ORCID: 0000-0001-5698-5505).}%
\thanks{G. Golan's ORCID is 0000-0002-1993-583X.}%
\thanks{All authors contributed equally to this work.}%
\thanks{Corresponding authors: Gilad Orr (e-mail: gilad.orr@ariel.ac.il) and Gady Golan (e-mail: gadygo@ariel.ac.il).}}
\title{A Modified Boost Converter Topology for Dynamic Characterization of
Hot Carrier and Trap Generation in GaN HEMTs}
\maketitle
\begin{abstract}
Modern microelectronic systems require long term operational stability,
necessitating precise reliability models to predict device lifecycles
and identify governing failure mechanisms. This is particularly critical
for high power GaN High-Electron-Mobility Transistors (HEMTs), where
reliability research has historically trailed behind low power digital
counterparts. This study introduces a novel application of a modified
boost converter circuit designed to investigate GaN failure mechanisms,
specifically targeting the determination of reliability factors for
the MTOL model. By utilizing a high duty cycle, the circuit stresses
the device at maximum rated voltages and currents with minimal input
requirements, accelerating hot carrier and trap generation without
immediate detrimental failure.

Experimental validation was conducted using an EPC 2038 GaN transistor
under a constant drain current of 400 mA and a duty cycle of 0.7.
The results confirmed that the increase in Drain-Source on-resistance
($R_{DS(on)}$) follows a logarithmic trend over time, consistent
with the EPC Phase 12 reliability model. While initial tests at 40V
did not successfully validate the longitudinal optical phonon scattering
energy ($\hbar\omega_{LO}$), but were reasonably acceptable, subsequent
stress tests at 70V and 100V yielded $\hbar\omega_{LO}$ values that
were successfully validated against existing theoretical and experimental
data. This methodology provides a robust framework for predicting
performance and lifetime across varying operational parameters in
modern power electronics.
\end{abstract}


\section{Introduction}

Over the last few years, the research in the field of GaN power transistors
has shown impressive advancements \cite{Meneghini2016PowerGaNDevices}.
Gallium nitride has a wide energy gap (3.4 eV) that allows operation
at high temperature. GaN transistors can remain fully functional well
above 300$^\circ$C, with excellent control of the channel current \cite{Gaska2014NovelAlInN}.
In addition, GaN has a breakdown field of 6 MV/cm \cite{orr2025gallium},
which is approximately ten times higher than that of silicon. For
the same breakdown voltage, GaN-based transistors are \textasciitilde 10
times shorter/thinner with respect to the silicon counterpart, and
this results in a significant reduction of the on-resistance of the
devices. GaN transistors with breakdown voltages higher than 1900V
have already been demonstrated \cite{Herbecq20141900V}, due to the
optimization of the buffer structure and/or the adoption of specific
methods for substrate removal. In AlGaN/GaN transistors, the channel
is formed at the hetero-interface between a larger band-gap material
(AlGaN) and GaN, due to the spontaneous and piezoelectric polarization
that occurs without any doping. For this reason, the mobility of the
channel is extremely high ($>2000cm^{2}/V\cdot s$ \cite{Chen2015RoomTemperatureMobility}),
resulting in current densities $>1,000,000\mu A/mm$.
Commercial $650V/60A$ devices can have a very low on-resistance $(<25m\Omega$,
\cite{GaNSystemsDatasheet}). this has a positive impact on the resistive
losses of the transistors, when they are used in switching mode power
converters and various pulsed high current applications \cite{orr2012safe}.
Finally, GaN transistors have a low Figure of Merit (FOM) $FOM=R_{DS(on)}\times Q_{g}$
(product of on-resistance and gate charge), smaller than a specific
value, $1nC\cdot\Omega$, and consequently very low switching loss.
Compared to the silicon counterparts, power converters based on GaN
will therefore be lighter (since they will require smaller heat sinks),
smaller (the lower switching losses enable to increase the switching
frequency, and to use smaller passives), and more efficient (thanks
to the lower resistive and switching losses). However, what about
reliability? In a switching-mode power converter (see the simplified
boost circuit in Figure 1), the transistor is continuously switched
from the off-state to the on-state and vice versa. When the transistor
is in off-state, a high voltage (maximum stress) is applied between
drain and substrate, drain and source, and between drain and gate.
\begin{figure}[h]
\begin{centering}


\begin{circuitikz}[american, scale=0.7, transform shape]
\draw[battery1={$V_{in}$},/tikz/circuitikz/bipoles/length=1.1cm](4.5,-4.5)to(4.5,-6.5);
\draw[L={},/tikz/circuitikz/bipoles/length=1.1cm](4.5,-4.5)to(8.5,-4.5);
\ctikzset{tripoles/mos style/arrows}
\draw node[nmos,scale=0.59,xscale=1,yscale=1,rotate=0](Q3) at (8.5,-5.5) {} node[anchor=west,scale=0.9] at (Q3.text){HEMT};
\node[anchor=west] at (6, -5.5) {Drive};
\draw[short](Q3.C)to(8.5,-5.0);
\draw[short](Q3.E)to(8.5,-6.0);
\draw[short={}](8.5,-4.5)to(8.5,-4.5);
\draw[short={}](8.5,-4.5)to(8.5,-5.0);
\draw[short={}](8.5,-6.0)to(8.5,-6.5);
\draw[short={}](8.5,-6.5)to(4.5,-6.5);
\draw[C={},/tikz/circuitikz/bipoles/length=1.1cm](12.0,-4.5)to(12.0,-6.5);
\draw[R={Load},/tikz/circuitikz/bipoles/length=1.1cm](13.5,-4.5)to(13.5,-6.5);
\draw[D={},/tikz/circuitikz/bipoles/length=1.1cm](8.5,-4.5)to(12.0,-4.5);
\draw[short={}](12.0,-4.5)to(13.5,-4.5);
\draw[short={}](13.5,-6.5)to(8.5,-6.5);
\draw node[circ] (S13) at (12.0, -6.5) {} node[anchor=south] at ([yshift=-0.6cm]S13.text){};
\draw node[circ] (S14) at (12.0, -4.5) {} node[anchor=south] at ([yshift=-0.6cm]S14.text){};
\draw node[circ] (S15) at (8.5, -4.5) {} node[anchor=south] at ([yshift=-0.6cm]S15.text){};
\draw node[circ] (S16) at (8.5, -6.5) {} node[anchor=south] at ([yshift=-0.6cm]S16.text){};
\draw[short={}](7.5,-5.5)to(8.0,-5.5);
\draw[short={}](7.5,-5.5)to(7.5,-5.5);
\draw[short={}](7.5,-5.5)to(6.0,-5.0);
\draw[short={}](6.0,-5.0)to(6.0,-6.0);
\draw[short={}](6.0,-6.0)to(7.5,-5.5);
\draw[short={}](4.5,-8.5)to(4.5,-13.5);
\draw[short={}](4.5,-13.5)to(13.5,-13.5);
\draw node[currarrow] (S24) at (13.5, -13.5) {} node[anchor=north] at ([yshift=-0.6cm]S24.text){};
\draw node[currarrow, rotate=90] (S25) at (4.5, -8.5) {} node[anchor=south] at ([yshift=-0.6cm]S25.text){};
\node[anchor=east] at (4.3, -10.5) {$V_{DS}$};
\node[anchor=north] at (5.4, -11) {$OFF$};
\node[anchor=north] at (5.4, -11.5) {$V_{GS}=0V$};
\node[anchor=north] at (5.4, -12) {$I_D=0A$};
\node[anchor=north] at (8.4, -11) {$ON$};
\node[anchor=north] at (8.4, -11.5) {$V_{GS}>5V$};
\node[anchor=north] at (8.4, -12) {$I_D>0A$};
\node[anchor=north] at (8.4, -13.6) {Time [s]};
\draw[short={}, ultra thick, red](4.5,-10.5)to(6.0,-10.5);
\draw[short={}, ultra thick, red](6.0,-10.5)to(7.5,-13.5);
\draw[short={}, ultra thick, red](7.5,-13.5)to(9.5,-13.5);
\draw[short={}, ultra thick, red](9.5,-13.5)to(11.0,-10.5);
\draw[short={}, ultra thick, red](11.0,-10.5)to(12.5,-10.5);
\node[anchor=west] at (12.7, -10.5) {$650V$};
\end{circuitikz}

\par\end{centering}
\caption{\label{fig:simplified-representation}(top) simplified representation
of a switching mode power converter (boost configuration); (bottom)
simplified representation of the different operating regimes of the
transistor in a boost converter.}

\end{figure}
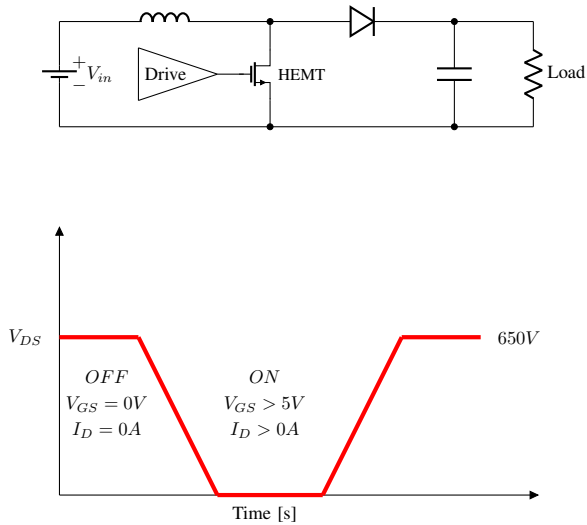
This may result in: 
\begin{itemize}
\item Buffer trapping processes that are responsible for the increase in
dynamic on-resistance; charge trapping may take place due to both
the ionization of acceptor states \cite{Moens2015ImpactOfBufferLeakage},
and to the injection of electrons from the substrate \cite{Bisi2014KineticsOfBuffer}. 
\item Injection of electrons at the gate-drain surface, which can be avoided
by using an optimized passivation layer. 
\item Time-dependent (permanent) degradation of the devices, consisting
of an increase in source-to-drain leakage \cite{Meneghini2014OFFStateDegradation},
or in the generation of short circuit paths between the gate and the
channel \cite{Rossetto2017FieldRelatedFailure}.
\end{itemize}
When the transistor is in the on-state, the gate is positively biased
at voltages higher than 5-6 V (we consider the case of a normally-off
device here). Normally-off GaN transistors can be fabricated via the
use of MIS/MOS structures with a partially-recessed gate \cite{Wu2015TimeDependentDielectric},
or by using a p-type gate stack \cite{Uemoto2007GateInjectionTransistor}.
Both technologies can show reliability issues when the devices are
stressed at high gate voltages \cite{tapajna2015investigation,Wu2013ComprehensiveInvestigation}.
Finally, when the transistor in Figure 1 switches from the off state
to the on-state, it crosses a semi-on condition, in which the voltage
and current on the drain may be simultaneously high. This \textquotedblleft hard-switching\textquotedblright{}
condition can originate from the discharge of the drain-source capacitance
of the transistor (when the HEMT is switched from off- to on-state
\cite{Joh2014CurrentCollapse,Bahl2016ProductLevelReliability}), and/or
from a poor optimization of the dead times of the switching events
(in a half-bridge circuit). The simultaneous presence of high current
and voltage on the drain may favor hot electron trapping/degradation
effects, which reduces the performance and/or reliability of the devices.

\section{TESTING METHOD \& SETUP CONFIGURATIONS}

\subsection{Boost Converter}

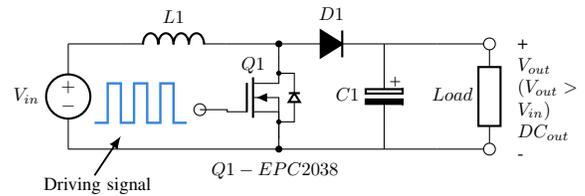
\begin{figure}[h]
\begin{centering}

\begin{tikzpicture}[american, scale=0.7, transform shape]
	\node[nigfetd, bodydiode](N1) at (7, 6.98){} node[anchor=south] at ([yshift=-0.44cm]N1.north){$Q1$};
	\draw (4, 8) to[american inductor, l={$L1$}] (6, 8);
	\draw (6, 8) -| (7, 7.75);
	\node[ocirc, xscale=2, yscale=2] at (5.5, 6.75){};
	\draw (5.5, 6.75) -| (6.02, 6.71);
	\path[draw={rgb,255:red,53;green,132;blue,228}, line width=0.9pt] (3.5, 6.5) -- (3.75, 6.5) -| (3.75, 7.25) -- (4, 7.25) -| (4, 6.5) -- (4.25, 6.5) -| (4.25, 7.25) -- (4.5, 7.25) -| (4.5, 6.5) -- (4.75, 6.5) -- (4.75, 7.25) -- (5, 7.25) -- (5, 6.5) -- (5.25, 6.5);
	\draw (3, 8) -- (4, 8);
	\draw (7, 8) to[full diode, /tikz/circuitikz/bipoles/length=1.02cm, l={$D1$}, name=D1] (9, 8);
	\draw (11, 8) to[european resistor, l_={$Load$}] (11, 6);
	\draw (9, 8) -- (10, 8);
	\draw (10, 6) -- (3, 6);
	\draw (7, 6.25) -| (7, 6);
	\node[circ] at (7, 8){};
	\node[circ] at (7, 6){};
	\draw (9, 8) to[ecapacitor, l_={$C1$}] (9, 6);
	\draw (10, 8) -- (11, 8);
	\draw (10, 6) -- (11, 6);
	\node[shape=rectangle, minimum width=-0.035cm, minimum height=-0.035cm] at (11.497, 8.106){} node[anchor=north west, align=justify, text width=-0.423cm, inner sep=6pt] at (11.297, 8.306){+ $V_{out}\\ (V_{out}>V_{in})$ $DC_{out}$\\-};
	\node[ocirc, xscale=2, yscale=2] at (11, 8){};
	\node[ocirc, xscale=2, yscale=2] at (11, 6){};
	\node[shape=rectangle, minimum width=2.965cm, minimum height=0.435cm] at (7, 5.735){} node[anchor=north west, align=left, text width=2.577cm, inner sep=6pt] at (5.5, 5.97){$Q1 - EPC2038$};
	\draw[line width=0.7pt, -latex] (3.5, 5.5) -- (4, 6.25);
	\node[shape=rectangle, minimum width=2.715cm, minimum height=0.215cm] at (3.693, 5.545){} node[anchor=north west, align=left, text width=2.327cm, inner sep=6pt] at (2.318, 5.67){Driving signal};
	\draw (3, 8) to[american voltage source, l_={$V_{in}$}] (3, 6);
\end{tikzpicture}
\par\end{centering}
\caption{\label{fig:Current-Path-with}Current Path with GaN Off}

\end{figure}

Switched-mode supplies can be used for many purposes, including DC-to-DC
converters. Often, a DC power source, such as a battery, may not provide
the required voltage by a given system. For example, the motors used
in driving electric automobiles require much higher voltages than
the ones directly accessible from batteries. Operating voltages of
modern electric cars are in the region of 500V. Even if banks of batteries
were used, the extra weight and space consumed would be too great
to be practical. The answer to this problem is to use fewer batteries
and to boost the available DC voltage to the required level by using
a boost converter. Another problem with batteries, large or small,
is that their output voltage varies as the available charge is used
up, and at some point, the battery voltage becomes too low to power
the circuit being supplied. However, if this low output level can
be boosted back up to a useful level again, by using a boost converter,
the life of the battery can be extended. The DC input to a boost converter
can be from many sources as well as batteries, such as rectified AC
from the mains supply, or DC from solar panels, fuel cells, dynamos
and DC generators. The boost converter is different to the Buck Converter in
that its output voltage is equal to, or greater than its input voltage.
Keeping in mind that power needs to be conserved, any voltage conversion
results in a change in current. Thus, increasing the voltage results
in a reduced current. Due to the high switching frequencies and currents,
boost converters stress their components considerably. In this paper,
we present reliability tests for GaN transistors in a boost converter
design. As demonstrated earlier, using a classic boost converter circuit,
we were able to stress test the GaN transistors to the full extent
of their maximum voltages and currents. This showcases that the circuits
are excellent for reliability testing. Figure \ref{fig:Current-Path-with}
illustrates the basic design of a Boost converter with GaN transistor.

\subsection{Boost Converter Operation}

\begin{figure}[h]
\begin{centering}

\begin{tikzpicture}[american, scale=0.7, transform shape]
	\draw (3, 8) to[american voltage source, l_={$V_{in}$}] (3, 6);
	\node[nigfetd, bodydiode](N1) at (7, 6.98){} node[anchor=south] at ([yshift=-0.44cm]N1.north){$Q1$};
	\draw (4, 8) to[american inductor, l={$L1$}] (6, 8);
	\draw (6, 8) -| (7, 7.75);
	\node[ocirc, xscale=2, yscale=2] at (5.5, 7){};
	\draw (3, 8) -- (4, 8);
	\draw (7, 8) to[full diode, /tikz/circuitikz/bipoles/length=1.02cm, l={$D1$}, name=D1] (9, 8);
	\draw (11, 8) to[european resistor, l_={$Load$}] (11, 6);
	\draw (9, 8) -- (10, 8);
	\draw (10, 6) -- (3, 6);
	\draw (7, 6.25) -| (7, 6);
	\node[circ] at (7, 8){};
	\node[circ] at (7, 6){};
	\draw (9, 8) to[ecapacitor, l_={$C1$}] (9, 6);
	\draw (10, 8) -- (11, 8);
	\draw (10, 6) -- (11, 6);
	\node[ocirc, xscale=2, yscale=2] at (11, 8){};
	\node[ocirc, xscale=2, yscale=2] at (11, 6){};
	\node[shape=rectangle, minimum width=3cm, minimum height=0.47cm] at (7.6, 5.735){} node[anchor=north west, align=left, text width=2.647cm, inner sep=5pt] at (6.1, 5.97){$Q1 \rightarrow ON$};
	\node[shape=rectangle, minimum width=2.715cm, minimum height=0.215cm] at (4.394, 5.749){} node[anchor=north west, align=left, text width=2.327cm, inner sep=6pt] at (3.019, 5.874){$V_{sig} \rightarrow High$};
	\path[draw={rgb,255:red,98;green,160;blue,234}, line width=1.1pt] (4.846, 6.142) -- (5.096, 6.142) -| (5.096, 6.892) -- (5.346, 6.892);
	\draw (5.5, 7) -- (6, 7) -| (6.02, 6.71);
	\draw[line width=0.7pt, -latex] (4, 5.75) -- (4.75, 6.25);
	
    \coordinate (C) at (4.3, 5.3);
    \draw[red, line width=1.5pt, ->, >=stealth] (C) ++(90:1cm) arc (260:-0:0.7cm);
    \node[shape=rectangle, minimum width=2cm, minimum height=0.465cm] at (4.408, 7.142){} node[anchor=north west, align=left, text width=2cm, inner sep=6pt] at (3.6, 7.32){\textcolor{rgb,255:red,237;green,51;blue,59}{$I_{in}>0$}};
\end{tikzpicture}
\par\end{centering}
\caption{\label{fig:Boost-Converter-Operation}Boost Converter Operation at
Switch On}

\end{figure}
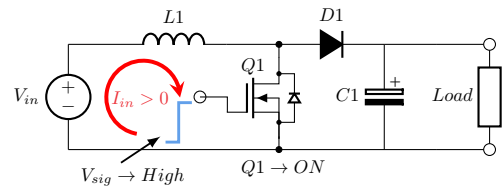

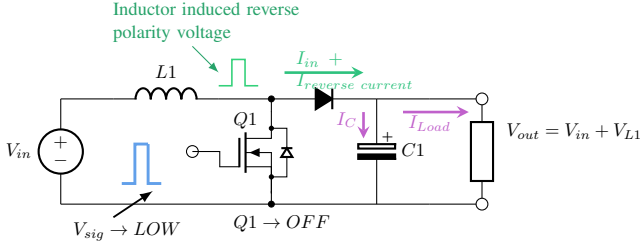
\begin{figure}[h]
\begin{centering}

\begin{tikzpicture} [american, scale=0.7, transform shape]
	\node[nigfetd, bodydiode](N1) at (7, 6.98){} node[anchor=south] at ([yshift=-0.44cm]N1.north){$Q1$};
	\draw (4, 8) to[american inductor, l={$L1$}] (6, 8);
	\draw (6, 8) -| (7, 7.75);
	\node[ocirc, xscale=2, yscale=2] at (5.5, 7){};
	\draw (3, 8) -- (4, 8);
	\draw (7, 8) to[full diode, /tikz/circuitikz/bipoles/length=0.784cm, name=D1] (9, 8);
	\draw (11, 8) to[european resistor] (11, 6);
	\draw (9, 8) -- (10, 8);
	\draw (10, 6) -- (3, 6);
	\draw (7, 6.25) -| (7, 6);
	\node[circ] at (7, 8){};
	\node[circ] at (7, 6){};
	\draw (9, 8) to[ecapacitor, l={$C1$}] (9, 6);
	\draw (10, 8) -- (11, 8);
	\draw (10, 6) -- (11, 6);
	\node[ocirc, xscale=2, yscale=2] at (11, 8){};
	\node[ocirc, xscale=2, yscale=2] at (11, 6){};
	\node[shape=rectangle, minimum width=3cm, minimum height=0.47cm] at (7.6, 5.735){} node[anchor=north west, align=left, text width=2.647cm, inner sep=5pt] at (6.1, 5.97){$Q1 \rightarrow OFF$};
	\node[shape=rectangle, minimum width=2.715cm, minimum height=0.215cm] at (4.394, 5.749){} node[anchor=north west, align=left, text width=2.327cm, inner sep=6pt] at (3.019, 5.874){$V_{sig} \rightarrow LOW$};
	\path[draw={rgb,255:red,98;green,160;blue,234}, line width=1.1pt] (4.159, 6.392) -- (4.408, 6.392) -| (4.408, 7.142) -- (4.658, 7.142);
	\path[draw={rgb,255:red,98;green,160;blue,234}, line width=1.1pt] (4.659, 7.142) -| (4.659, 6.392) -- (4.909, 6.392);
	\draw (5.5, 7) -- (6, 7) -| (6.02, 6.71);
	\draw[line width=0.7pt, -latex] (4, 5.75) -- (4.75, 6.25);
	\node[shape=rectangle, minimum width=1.34cm, minimum height=0.465cm] at (8.709, 7.718){} node[anchor=north west, align=left, text width=0.952cm, inner sep=6pt] at (8.022, 7.968){\textcolor{rgb,255:red,192;green,97;blue,203}{$I_{C}$}};
	\draw (3, 8) to[american voltage source, l_={$V_{in}$}] (3, 6);
	\path[draw={rgb,255:red,51;green,209;blue,122}, line width=0.6pt] (6, 8.25) -- (6.25, 8.25) -| (6.25, 8.75) -- (6.5, 8.75) -| (6.5, 8.25) -- (6.75, 8.25);
	\node[shape=rectangle, minimum width=4.215cm, minimum height=0.465cm] at (5.902, 9.794){} node[anchor=north west, align=left, text width=3.827cm, inner sep=6pt] at (3.777, 10.044){\textcolor{rgb,255:red,38;green,162;blue,105}{Inductor induced reverse polarity voltage}};
	\path[draw={rgb,255:red,38;green,162;blue,105}, -latex] (5.5, 9) -- (6, 8.5);
	\path[draw={rgb,255:red,192;green,97;blue,203}, line width=1pt, -stealth] (8.75, 7.75) -| (8.75, 7.25);
	\path[draw={rgb,255:red,192;green,97;blue,203}, line width=1pt, -stealth] (9.5, 7.75) -- (10.75, 7.75);
	\node[shape=rectangle, minimum width=1.215cm, minimum height=0.465cm] at (10.011, 7.653){} node[anchor=north west, align=left, text width=0.827cm, inner sep=6pt] at (9.386, 7.903){\textcolor{rgb,255:red,192;green,97;blue,203}{$I_{Load}$}};
	\path[draw={rgb,255:red,46;green,194;blue,126}, line width=1pt, -stealth] (7.25, 8.5) -- (8.75, 8.5);
	\node[shape=rectangle, minimum width=2.965cm, minimum height=0.715cm] at (8.75, 8.75){} node[anchor=north west, align=left, text width=2.577cm, inner sep=6pt] at (7.25, 9.125){\textcolor{rgb,255:red,46;green,194;blue,126}{$I_{in}+I_{reverse\ current}$}};
	\node[shape=rectangle, minimum width=2.965cm, minimum height=0.715cm] at (12.75, 7.375){} node[anchor=north west, align=left, text width=2.577cm, inner sep=6pt] at (11.25, 7.75){$V_{out}=V_{in}+V_{L1}$};
\end{tikzpicture}
\par\end{centering}
\caption{\label{fig:Current-Path-with-1}Current Path with GaN Off}

\end{figure}

\begin{figure}[h]
\begin{centering}

\begin{tikzpicture}[american, scale=0.7, transform shape]
	\node[nigfetd, bodydiode](N1) at (7, 6.98){} node[anchor=south] at ([yshift=-0.44cm]N1.north){$Q1$};
	\draw (4, 8) to[american inductor, l={$L1$}] (6, 8);
	\draw (6, 8) -| (7, 7.75);
	\node[ocirc, xscale=2, yscale=2] at (5.5, 7){};
	\draw (3, 8) -- (4, 8);
	\draw (7, 8) to[full diode, /tikz/circuitikz/bipoles/length=0.784cm, l={$D1$}, label distance=-0.06cm, name=D1] (9, 8);
	\draw (11, 8) to[european resistor] (11, 6);
	\draw (9, 8) -- (10, 8);
	\draw (10, 6) -- (3, 6);
	\draw (7, 6.25) -| (7, 6);
	\node[circ] at (7, 8){};
	\node[circ] at (7, 6){};
	\draw (9, 8) to[ecapacitor, l_={$C1$}] (9, 6);
	\draw (10, 8) -- (11, 8);
	\draw (10, 6) -- (11, 6);
	\node[ocirc, xscale=2, yscale=2] at (11, 8){};
	\node[ocirc, xscale=2, yscale=2] at (11, 6){};
	\node[shape=rectangle, minimum width=3cm, minimum height=0.47cm] at (7.6, 5.735){} node[anchor=north west, align=left, text width=2.647cm, inner sep=5pt] at (6.1, 5.97){$Q1 \rightarrow HIGH$};
	\node[shape=rectangle, minimum width=2.715cm, minimum height=0.215cm] at (4.75, 5.375){} node[anchor=north west, align=left, text width=2.327cm, inner sep=6pt] at (3.375, 5.5){$V_{sig} \rightarrow LOW$};
	\path[draw={rgb,255:red,98;green,160;blue,234}, line width=1.1pt] (4.402, 6.094) -- (4.652, 6.094) -| (4.652, 6.844) -- (4.902, 6.844);
	\path[draw={rgb,255:red,98;green,160;blue,234}, line width=1.1pt] (4.902, 6.844) -| (4.902, 6.094) -- (5.152, 6.094);
	\path[draw={rgb,255:red,98;green,160;blue,234}, line width=1pt] (5.152, 6.094) -| (5.152, 6.844) -- (5.402, 6.844);
	\draw (5.5, 7) -- (6, 7) -| (6.02, 6.71);
	\draw (3, 8) to[american voltage source, l_={$V_{in}$}] (3, 6);
	\node[shape=rectangle, minimum width=2.965cm, minimum height=0.715cm] at (12.75, 7.375){} node[anchor=north west, align=left, text width=2.577cm, inner sep=6pt] at (11.25, 7.75){$V_{out}=V_{C1}$};
	\draw[line width=1pt, -latex] (4.5, 5.5) -- (5.25, 6.5);
	\node[shape=rectangle, minimum width=2cm, minimum height=0.465cm] at (4.408, 7.142){} node[anchor=north west, align=left, text width=2cm, inner sep=6pt] at (3.9, 7.32){\textcolor{rgb,255:red,237;green,51;blue,59}{$I_{in}$}};
	\node[shape=rectangle, minimum width=2cm, minimum height=0.465cm] at (4.408, 7.142){} node[anchor=north west, align=left, text width=2cm, inner sep=6pt] at (9.4, 7.32){\textcolor{rgb,255:red,237;green,51;blue,59}{$I_{out}$}};
    \coordinate (C) at (4.3, 5.3);
    \draw[red, line width=1.5pt, ->, >=stealth] (C) ++(90:1cm) arc (260:-0:0.7cm);
    \coordinate (C) at (9.3, 6.4);
    \draw[red, line width=1.5pt, ->, >=stealth] (C) ++(90:1cm) arc (150:-70:0.7cm);
\end{tikzpicture}
\par\end{centering}
\caption{\label{fig:Current-Path-with-Gan-On}Current Path with GaN On}

\end{figure}
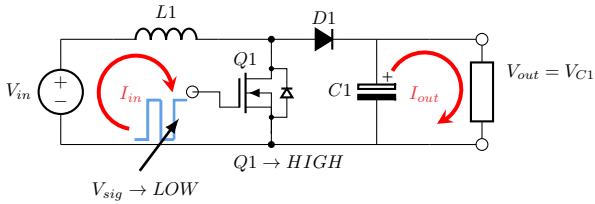

Figure \ref{fig:Boost-Converter-Operation} illustrates the circuit
action during the initial high period of the high frequency square
wave applied to the GaN gate at start up. During this time GaN conducts,
placing a short circuit from the right-hand side of L1 to the negative
input supply terminal. Therefore a current flows between the positive
and negative supply terminals through L1, which stores energy in its
magnetic field. There is virtually no current in the remainder of
the circuit as the combination of D1, C1 and the load represent a
much higher impedance than the path directly through the heavily conducting
GaN transistor.
Figure \ref{fig:Current-Path-with-1} shows the current path during
the low period of the switching square wave cycle. As the GaN transistor
is rapidly turned off the sudden drop in current causes L1 to produce
a back e.m.f. in the opposite polarity to the voltage across L1 during
the on period, to keep current flowing. This results in two voltages,
the supply voltage $V_{in}$ and the back e.m.f.($V_{L}$) across
L1 in series with each other. This higher voltage ($V_{in}+V_{L}$),
now that there is no current path through the GaN transistor, forward
biases D1. The resulting current through D1 charges up C1 to $V_{in}+V_{L}$
minus the small forward voltage drop across D1 and also supplies the
load.
Figure \ref{fig:Current-Path-with-Gan-On} shows the circuit action
during GaN on periods after the initial start-up. Each time the GaN
conducts, the cathode of D1 is more positive than its anode, due to
the charge on C1. D1 is therefore turned off so the output of the
circuit is isolated from the input, however the load continues to
be supplied with $V_{in}+V_{L}$ from the charge on C1. Although the
charge C1 drains away through the load during this period, C1 is recharged
each time the GaN switches off, so maintaining an almost steady output
voltage across the load. The theoretical DC output voltage is determined
by the input voltage ($V_{in}$) divided by 1 minus the duty cycle
(D) of the switching waveform, which will be some figure between 0
and 1 (corresponding to 0 to 100\%) and therefore can be determined
using the following formula:

$V_{out}=\frac{V_{in}}{1-D}$

\subsubsection{Boost convertor based on the EPC 2038 GaN transistor}

The boost converter that we are showing in this paper is based on
GaN n-channel transistor (EPC 2038) which used as switch, similar
to the traditional Boost converter circuit with one main difference:
the output is connected to a high-voltage supply instead of a regular
load (Figure \ref{fig:Electrical-diagram-of}). We present the GaN
EPC 2038 characteristics in Table \ref{tab:-EPC-2038}.

\begin{table}[h]
\begin{centering}
\begin{tabular}{cccc}
\hline 
Parameter & Pulsed Input & Continuous Input & Unit\tabularnewline
\hline 
$V_{DS}(max)$ & $120$ & $100$ & V\tabularnewline
$I_{D}(max)$ & $0.5$ & $0.5$ & A\tabularnewline
$V_{GS}(max)$ & $6$ & $6$ & V\tabularnewline
$V_{GS}(min)$ & $-5$ & $-5$ & V\tabularnewline
$T_{J}$ & $-40:150$ & $-40:150$ & $^{\circ}C$\tabularnewline
$R_{DS(ON)}$ & $3.3$ & $3.3$ & $\Omega$\tabularnewline
\hline 
\end{tabular}
\par\end{centering}
\caption{\label{tab:-EPC-2038}$GaN$ EPC 2038 characteristics}

\end{table}

\subsubsection*{a. Objectives of the boost converter circuit}

A new boost converter test vehicle for High Voltage Reverse Bias (HVRB)
where it will allow us to stress the devices at full rated voltage
and maximum expected current. The test will be performed using a modified
Boost-converter circuit where a high duty cycle will allow the maximum
voltage output at the drain using a minimal input voltage. A diode
is connected to the drain so that the maximum voltage is seen across
the EPC device while the gate is biased to 0V (Off state). This test
will be performed in a matrix of high voltage and temperature in order
to find an expression for time to fail versus Voltage and Temperature
(giving us an Arrhenius relation too). All our testing will include
careful monitoring of the gate voltage since there is a known failure
mechanism relating to leakage in the gate junction.

\subsubsection*{b. Concept and approach of boost converter circuit}

The principle of operation of the circuit in Figure \ref{fig:Electrical-diagram-of}
is similar to the regular boost converter that was presented in the
previous section, as the following: the current through the inductor
$L_{drain}\,(10\mu H)$ increases linearly under constant voltage,
while the average voltage of the inductor is zero. Hence, the applied
voltage ($V_{in}$) is equal to the average voltage applied to the
Drain. When the Pulse Generation (PG) is on, the current increases
across the inductor and the voltage on the inductor is positive, and
while the PG is off the voltage on $V_{DS}$ increases rapidly across
the device under test and the inductor voltage turns negative while
the current decreases. The energy from the inductor goes into the
capacitor and the capacitance of the device and when the voltage exceeds
the maximum output voltage, the spikes flow through the diode to the
output voltage supply. To test the logic operation of the circuit
we employed the DUT with NFET GaN (2038). All the parameters of the
other components of the circuit kept identical for the entire circuits
and experiments of the test: $L=10\lyxmathsym{\textmu}H$, $C_{in}=100pF$,
$Vout=100V$ and $C_{out}=25pF$. The average source voltage ($V_{in}$)
is the voltage across the inductor plus the voltage across the device.
Since the voltage across an inductor is by definition equal to zero,
then we can say that the average voltage across the device is equal
to the average voltage at the input. The Duty cycle, D, describes
the ON-time over the total ON plus OFF time. Hence: 
\begin{equation}
\bar{V}_{in}=I(on)\times R_{DS(on)}\times D+V_{max}\cdot\frac{(1-D)}{2}\label{eq:Vin_average}
\end{equation}
where we assume that the rise and fall times of the output voltage
is triangular. With an output diode, there may be more time at $V_{max}$,
so we may need to modify a factor of 2. Nonetheless, we see that most
of the on-time relates to the voltage drop across the transistor when
it is ON. Then, by re-arranging Equation \ref{eq:Vin_average} we
can find an expression for $R_{DS(ON)}$: 
\begin{equation}
R_{DS(on)}=\frac{(V_{in}\lyxmathsym{\textendash}V_{max}\frac{(1-D)}{2})}{\bar{I}\cdot D}\label{eq:Rds(on)}
\end{equation}
\begin{figure}[h]
\begin{centering}

\begin{tikzpicture}[american, scale=0.6, transform shape]
	\node[nigfetd, bodydiode](Q1) at (8, 4.23){} node[anchor=south west] at ([xshift=-0.33cm, yshift=-0.33cm]Q1.north east){$EPC2038$};
	\draw (3.5, 8) to[american inductor, l={$100p$}] (5.5, 8);
	\draw (9, 6) to[full diode, /tikz/circuitikz/bipoles/length=0.784cm, l={$MBRS1100$}, label distance=-0.06cm, name=D1] (11, 6);
	\draw (3, 8) to[american voltage source, l_={$V_{in}=10V$}, label distance=0.25cm] (3, 6);
	\draw (6, 7.5) to[capacitor, l={$10\mu F$}] (6, 6.5);
	\draw (6, 6.5) -| (6, 6) |- (3, 6);
	\draw (6, 7.5) -| (6, 8);
	\draw (3, 8) -- (3.5, 8);
	\draw (5.5, 8) -- (6, 8);
	\draw (6, 8) to[american inductor, l={$100p$}] (8, 8);
	\draw (8, 8) to[american inductor, l={$L_{drain}=10\mu H$}] (8, 6);
	\draw (9, 6) -- (8, 6);
	\draw (8, 5) -| (8, 6);
	\node[circ] at (8, 6){};
	\node[circ] at (6, 8){};
	\draw (12, 6) to[american inductor, l={$100p$}] (12, 4.5);
	\draw (11, 6) -- (12, 6);
	\draw (12, 4.5) to[american voltage source, l={$100V$}] (12, 3);
	\draw (5, 3.75) to[square voltage source, l={$V_{pulse}$}] (5, 2.25);
	\draw (5, 4) to[american inductor, l={$0.001p$}] (7.02, 3.96);
	\draw (8, 3.46) to[american inductor, l={$0.1n$}] (8, 1.75);
	\draw (3, 6) -| (3, 1.75) -- (8, 1.75) -| (12, 3);
	\draw (5, 2.5) -| (5, 1.75);
	\node[circ] at (5, 1.75){};
	\node[circ] at (3, 6){};
	\node[circ] at (8, 1.75){};
	\node[ground] at (7, 1.75){};
	\node[circ] at (7, 1.75){};
	\draw (5, 4) -| (5, 3.75);
\end{tikzpicture}
\par\end{centering}
\caption{\label{fig:Electrical-diagram-of}Electrical diagram of Boost Converter
circuit}

\end{figure}
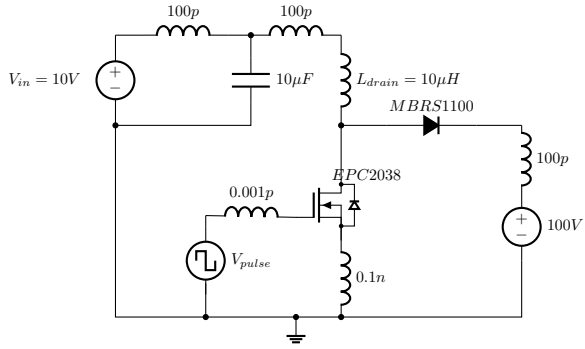
\begin{figure}[h]
\begin{centering}
\includegraphics[scale=0.6]{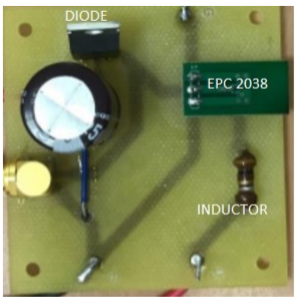}
\par\end{centering}
\caption{GaN Based boost converter assembled board}

\end{figure}
\begin{figure}[h]
\begin{centering}
\includegraphics[scale=0.6]{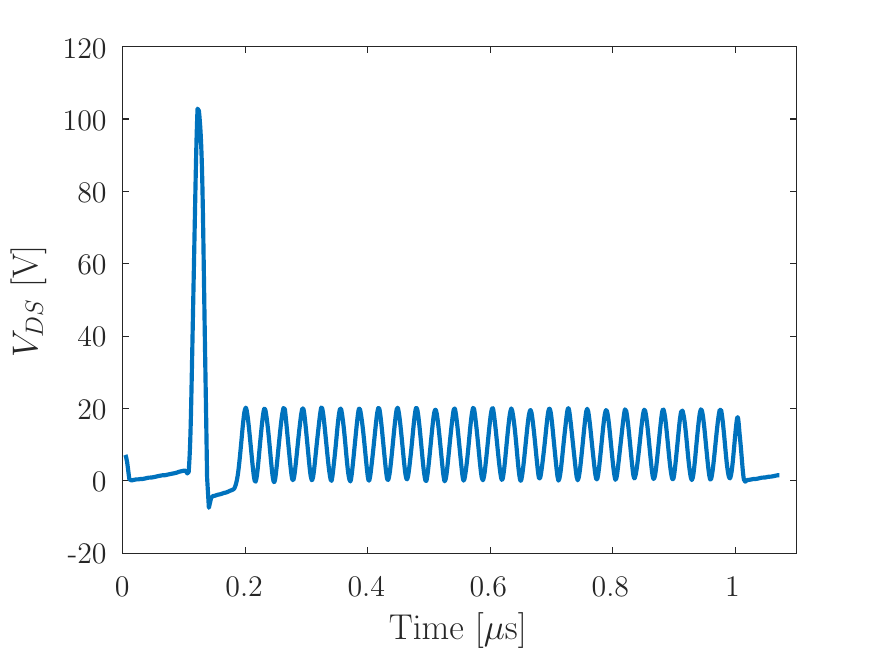}
\par\end{centering}
\caption{Simulation result which presents the voltage of $V_{DS}$ versus time,
$V_{DS}$ is reaching 100V}

\end{figure}

\begin{figure}[h]
\begin{centering}
\includegraphics[scale=0.6]{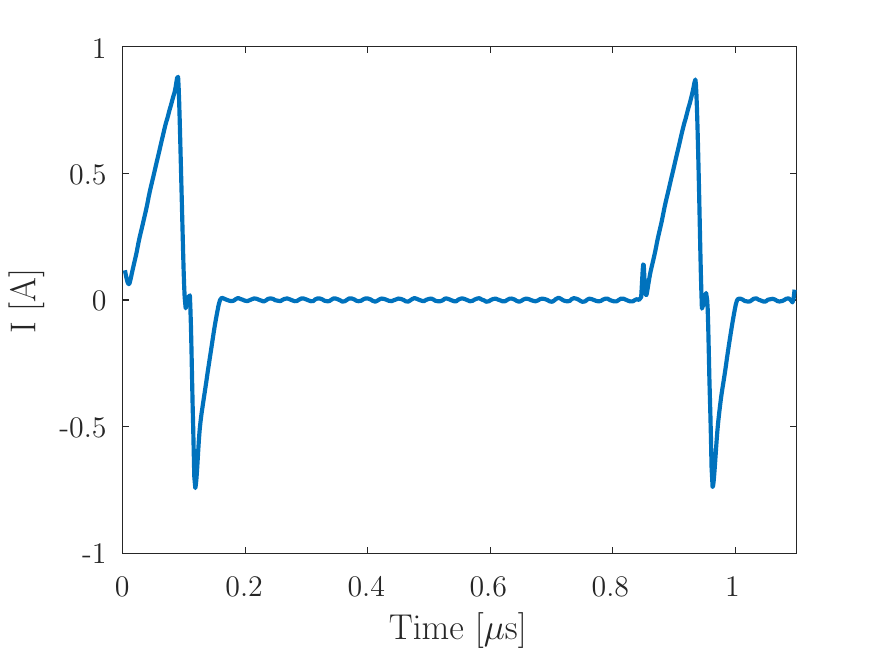}
\par\end{centering}
\caption{Simulation result which presents the current versus time, the peak
current is reaching almost 1A}

\end{figure}

\section{$GaN$ DEGRADATION MECHANISMS}

As the device design and material processing technology for AlGaN/GaN
HEMTs has matured over the recent years, several failure mechanisms
that limited device lifetime have been addressed and improved. These
mechanisms can be grouped together into three main categories that
effects lifetime: Contact degradation, hot electron effect (HCI),
and inverse piezoelectric effect. Both Schottky and Ohmic contacts
have shown excellent stability below 300 $^\circ$C \cite{golan2018improved}.
Piazza et al. \cite{Gao2011RoleOfOxygen} have reported an increase
in contact resistance and passivation cracking due to Ga out-diffusion
and Au inter-diffusion after a 100 h thermal storage test stress at
340 $^\circ$C. Nickel based Schottky contacts were found to form nickel nitrides
on GaN at annealing temperatures as low as 200 $^\circ$C, resulting in a
significant decrease in Schottky barrier height \cite{Gao2014ImpactOfWaterAssisted,Ambacher2002PyroelectricProperties}.
The observed current collapse and gate lag in AlGaN/GaN HEMTs under
high voltage and high current operation have been attributed to hot
electrons. These are electrons that have been accelerated in a large
electric field, resulting in very high kinetic energy, which can cause
trap formation. Creation of traps can occur in both the AlGaN layer
and the buffer, leading to reversible degradation of transconductance
and saturated drain current \cite{Gaska2014NovelAlInN,Uemoto2007GateInjectionTransistor}.
In this paper we are focusing on both hot Carriers and Trap Generation
and Contact Degradation as we are trying to reach the ideal conditions
for both mechanisms using the boost converter circuit as we mentioned
before.

\subsection{Hot -Carriers and Trap Generation}

Permanent device degradation after high VDG stress under on-state
conditions has been attributed to the presence of hot electrons. In
GaAs-based devices, hot electrons generate holes which are accumulated
near the gate resulting in a negative shift of VT \cite{Marcon2010ComprehensiveReliability,Suehle2002UltrathinGateOxide,Rossetto2016TimeDependentFailure}.
Typically, IG is used to derive field-acceleration laws for failure.
Impact ionization, however, is negligible in GaN HEMTs. This is due
to the fact that tunneling injection dominates gate current, preventing
gate current from being used as an indicator for hot electron degradation
{[}1,4{]}. However, these hot electrons likely lead to trap generation
at the AlGaN/GaN interface and/or at the passivation GaN cap interface.
As in GaAs and InP based HEMTs, traps lead to an increase in the depletion
region between the gate and the drain, ultimately resulting in an
increase in drain resistance and subsequently a decrease in saturated
drain-source current. Comparatively, under reverse bias or so-called
off-state conditions the degradation is greatly reduced due to the
reduction of electrons present in the channel. Sozza et al. {[}8{]}
showed that GaN/AlGaN/GaN HEMTs that underwent a 3000 h on-state stress
resulted in an increase in surface traps with an activation energy
of about 0.55 eV. On the other hand, devices stressed under off-state
conditions saw a very small increase in traps. Meneghesso et al. have
employed electroluminescence (EL) to study the effect of hot-carriers
and its dependence on stress conditions {[}2{]}. Uniform EL emission
was observed along the channel for devices stressed at $V_{GS}=0V$
and $V_{DS}=20V$, which is due to hot electrons. However, there is
no presence of hot spots or current crowding. On the other hand, under
OFF state conditions with $V_{GS}=-6V$ and $V_{DS}=20V$ (resulting
in a $V_{GD}=-26V$), the EL emission from the channel is not uniform.
These hot spots may be due to injection of electrons from the gate
into the channel. Due to the high bias conditions, the electrons acquire
enough energy to give rise to photon emission.

\subsection{Contact Degradation}

Contact degradation and gate sinking are significant degradation mechanisms
at elevated temperatures in GaAs and InP based HEMTs. This has not
yet proven to be a significant issue with AlGaN/GaN HEMTs at temperatures
below 400 $^\circ$C for Pt/Au Schottky contacts and Ti/Al/Pt/Au annealed
Ohmic contacts \cite{Meneghini2016PowerGaNDevices,Uemoto2007GateInjectionTransistor,Rossetto2016StudyOfTheStability,Meneghini2015ExtensiveInvestigation,Degraeve1998NewInsights,Stathis1999PercolationModels,Meneghini2016NormallyOffGaN,inberg2001material,axelevitch2014application}.
An increase in Schottky barrier height was observed for Ni/Au Schottky
contacts after DC stress at elevated junction temperatures (200 $^\circ$C)
\cite{Meneghini2016PowerGaNDevices,Uemoto2007GateInjectionTransistor,Rossetto2015DemonstrationOfField,Chen2012HBMEsdRobustness,Shankar2016UniqueEsdBehavior,Kuzmik2004ElectricalOverstress,axelevitch2014application}.
This was due to consumption of an interfacial layer between the Schottky
contact and the AlGaN layer. Though the resulting positive shift in
the Schottky barrier height, and thus the pinch-off voltage, is ideal,
the subsequent change in IDSS is not favorable. Unstressed devices
were subjected to an anneal after the Schottky contact was deposited
in order to decrease the interfacial layer between the gate and semiconductor.
Devices that underwent the gate anneal showed 50\% less degradation
during a 24 h stress test as opposed to devices that did not receive
a gate anneal \cite{Meneghini2016PowerGaNDevices,Uemoto2007GateInjectionTransistor}.
Thermal storage tests up to 2000 h on Ti/Al/Ni/Au ohmic contacts at
and above 290 $^\circ$C showed an increase in contact resistance as well
as surface roughness due to growth of Au-rich grains that ultimately
led to cracks in passivation \cite{Rossetto2015DemonstrationOfField,Chen2012HBMEsdRobustness,Shankar2016UniqueEsdBehavior,Khan1994CurrentVoltageCharacteristic,Vetury2001ImpactOfSurfaceStates,Hu2001Si3N4AlGaN,Leirer2001RFCharacterization,Kuzmik2004ElectricalOverstress,golan2000novel}.
The two primary degradation mechanisms were determined to be Au inter-diffusion
within the metal layers and Ga out-diffusion from the semiconductor
into the metallic compounds. Similar degradation was observed after
DC stress tests that resulted in junction temperatures equivalent
to the thermal storage tests. Due to the high-power capability of
AlGaN/GaN HEMTs, proper temperature management is needed in order
to optimize device performance under high current and high voltage
operation \cite{Kim2003EffectsOfSiNPassivation,Ibbetson2000PolarizationEffects,Gao2014ImpactOfWaterAssisted,Ambacher2002PyroelectricProperties,Joh2008CriticalVoltage,valizadeh2005effects}.
Self-heating of devices can ultimately result in poor device performance
through contact degradation. Reliability of contacts is highly dependent
upon both metal schemes as well as processing during fabrication,
EPC's eGaN transistors that were used in this research do not use
a Schottky gate and therefore are not vulnerable to this whole contact
mechanism.

\begin{figure}[h]
\begin{centering}
\includegraphics[scale=0.4]{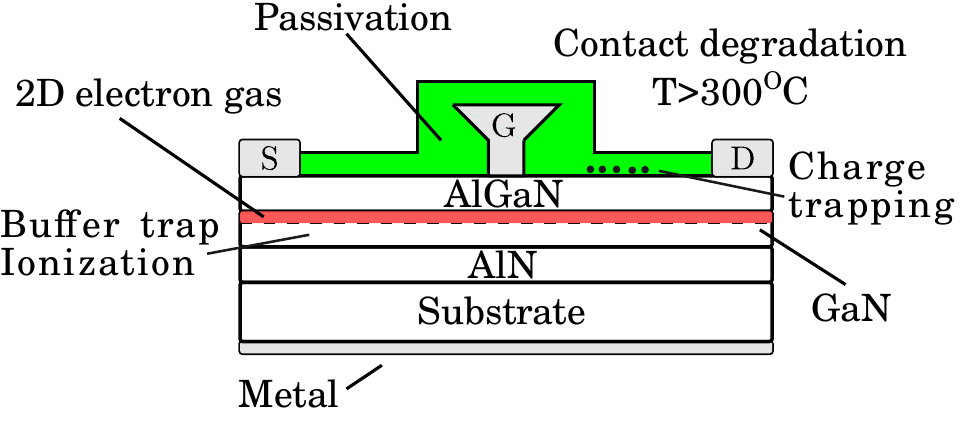}
\par\end{centering}
\caption{\label{fig:Schematic-of-degradation}Schematic of degradation mechanisms
in AlGaN/GaN HEMTs}

\end{figure}

Figure \ref{fig:Schematic-of-degradation} summarizes the reported
degradation mechanisms in AlGaN/GaN HEMTs during electrical stressing
at temperatures up to the typical operating temperature.

\section{EPCs $R_{DS(on)}$ model for GaN transistors}

In power electronics $R_{DS(on)}$ (Drain source on resistance) is
the most important metric required for device efficiency. Particularly,
in GaN transistors it is critical as it defines the tradeoff between
power delivery and dissipated power at a given size. $R_{DS(on)}$
depends both on time and temperature, creating a cycle in which an
increase in resistance leads to more dissipated power and hence to
an additional increase in resistance. This cycle can result in a thermal
runaway and hence a model for the evolution of $R_{DS(on)}$ with
time and temperature is crucial.

\subsection{A first principles mathematical model}

EPC has developed a first principles mathematical model to describe
the dynamic $R_{DS(on)}$ effect in eGaN FETs from the basic physics
of hot carrier scattering into surface traps. The model is published
in Appendix B of their Reliability report Phase 12 \cite{epc2020}. 

The published model describes the normalized $R_{DS(on)}$ as a function
of time temperature and drain voltage:


\begin{multline}
\frac{\Delta R}{R}=a+b\log_{10}\left(1+\exp\left(\frac{V_{DS}-V_{FD}}{\alpha}\right)\right)\sqrt{T}\\
\times\exp\left(\frac{\hbar\omega_{LO}}{k_{B}T}\right)\log(t)\label{eq:EPCs-GaN-RDSON-Model}
\end{multline}


\textbf{Independent Variables:}

$V_{DS}$ = Drain voltage (V)

$T$ = Device temperature (K)

t = Time (min)

\textbf{Parameters}:

a = 0.00 (unitless)

b = $2.0E-5\,(E^{-1/2})$

$\hbar\omega_{LO}=92$ meV (See the appendix for a detailed discussion
regarding the value of the \emph{longitudinal optical phonon scattering})

$V_{FD}$= 100V (appropriate for Gen5 100V products only)

a = 10 {[}V{]}

The model involves five device-dependent parameters. The values of
the parameters shown in Table \ref{tab:-EPC-2038} are appropriate
for an EPC2038 (DUT in this work) or other 100 V, 5'th Generation,
FETs listed in the appendix. The general form of this equation applies
to all eGaN FETs.

The EPC group also mentioned the following conclusions: 
\begin{itemize}
\item $R_{DS(on)}$ grows with time as log(t) 
\item The slope of $R_{DS(on)}$ over time has a negative temperature coefficient
(i.e., lower slope at higher temperature) 
\item Switching frequency does not affect the slope, but causes a small
vertical offset 
\item Switching current does not affect the slope, but causes a small vertical
offset 
\item Negligible difference between inductive and resistive hard switching.
\end{itemize}

\subsection{Time evolution of $R_{DS(on)}$ - validation}

If we slightly manipulate the model described in Equation~\ref{eq:EPCs-GaN-RDSON-Model}
we obtain the increase of $R_{DS(on)}$ with time, that is:\begin{equation} \begin{aligned} \Delta R_{DS(on)}(t) &= \Biggl[ a + \underbrace{b\log_{10}\left(1+e^{\frac{V_{DS}-V_{FD}}{\alpha}}\right) \sqrt{T}e^{\frac{\hbar\omega_{LO}}{k_{B}T}}}_{\text{slope}} \\ &\quad \cdot \log(t) \Biggr] \cdot R_{DS(on)}(0)\label{eq:R_DSon-evolution} \end{aligned} \end{equation}

which is a linear equation with a slope of 
\begin{equation}
s=b\log_{10}\left(1+\exp\left(\frac{V_{DS}-V_{FD}}{\alpha}\right)\right)\sqrt{T}\exp\left(\frac{\hbar\omega_{LO}}{k_{B}T}\right)\label{eq:slope}
\end{equation}

with $s$ designating the slope which is temperature dependent. This
point is important as we will now develop the means with which we
will try validate our experimental results and technique. A key physical
parameter in our validation process is the \emph{longitudinal optical
phonon scattering} (LO) - $\hbar\omega_{LO}$ above the temperature
of $250K$ which is considered to be around $92\,meV$ in GaN \cite{fang2019electron},
based on first principles band structure calculations. Thus we can
extract this parameter from Equation \ref{eq:slope}:
\begin{align}
\hbar\omega_{LO}(s,T,V_{DS}) & =\ln\left(\frac{s}{b\cdot\log_{10}(1+\exp\left(\frac{V_{DS}-V_{FD}}{\alpha}\right)\cdot\sqrt{T}}\right)\label{eq:longitudinal_optical_phonon}\\
 & \cdot k_{B}T\nonumber 
\end{align}
while the temperature $T$ and the drain source strain voltages are
independent variables, the slope $s$ depends on both that is $s(T,V_{DS})$.
Hence, when using error propagation analysis for error (uncertainty)
estimation we have to take into account that the error sources are
dependent and their absolute values need to be summed up. For simplicity,
we assume that $V_{DS}$ is stiff and does not contribute to the error.
$s$ is obtained from the time based experimentally measured $R_{DS(on)}$
being the major contributor to the error. Therefore, the uncertainty
in $\hbar\omega_{LO}$ is:

\begin{equation} \begin{aligned} \delta\hbar\omega_{LO}(s,T) &= \frac{\partial\hbar\omega}{\partial s}\delta s + \frac{\partial\hbar\omega}{\partial T}\delta T \\ &= \frac{k_{B}T}{s}\delta s + \Biggl[ \ln \left( \frac{s}{b\log_{10}(1+e^{\frac{V_{DS}-V_{FD}}{\alpha}})\sqrt{T}} \right) \\ &\quad - \frac{1}{2} \Biggr] \cdot k_{B}\delta T\label{eq:uncertainty} \end{aligned} \end{equation}As
the time in our measurements was measured in hours both Equations
\ref{eq:longitudinal_optical_phonon}\&\ref{eq:uncertainty} were
divided by $\log_{10}(60)$. The uncertainty we have just derived
in Equation \ref{eq:uncertainty} will assist us in qualifying the
suggested test circuit.


\section{RESULTS OF THE EXPERIMENTAL TESTS}

In this section we will present the preliminary results of the very
first tests that were done using the boost converter circuit. 

\subsection{Boost converter HVRB continues test}

Boost convertor circuit was activated with a constant drain current
source of 400 mA, while VD voltage was continuously measured during
the test, the analysis sequence of every experiment was a result of
monitoring VD voltage continuously the test was performed in four
different voltages gradually to make sure that the system is working
fine. The purpose of this test is to show that the charge trapping
mechanism responsible for a long-term increase of $R_{DS(on)}$ follows
a log (time) trend. The calculation of the $R_{DS(on)}$ was based
on Equation \ref{eq:Rds(on)}

In this case the $V_{in}$ is equal to the measured $V_{DS}$, $V_{max}$
was measured from the oscilloscope, and it varies from experiment
to experiment, $\bar{I}$ is the constant drain current source that
was mentioned before and were limited to 400 mA here. Where D is the
duty cycle, which is set to be 0.7 for this experiment. For example,
for the test that we will present next, the test was performed with
40V stress which led to the $V_{max}$ of 60v which measured from
the scope, the first sample of the $V_{DS}$ was equal to 9.95v, let's
put all the values in the Equation\ref{eq:Rds(on)}:

\begin{align*}
R_{DS}(on) & =\frac{\frac{9.95\text{\textendash}60(1-0.7)}{2}}{0.4\times0.7}=\\
R_{DS}(on) & =\frac{0.95}{0.28}=3.39\Omega
\end{align*}

As we can see, the result of the calculated $R_{DS}$ on is very close
to the $R_{DS}$ on of the EPC 2038 transistor which is 3.33 \textOmega,
in the following graphs we will show the results of $R_{DS}$ on vs
time and normalized $R_{DS}$ on vs time based on the equation and
the calculations that were shown above.

\subsubsection{Test with 40V stress, 400mA}

This test was performed with a constant drain current of 400 mA and
constant output voltage of 40v ,$V_{max}$ reached 60 v ; below in
Figure \ref{fig11} we present the change of the $R_{DS(on)}$ during
this test which affected by the change of $V_{DS}$:
\begin{figure}[h]
\begin{centering}
\includegraphics[scale=0.6]{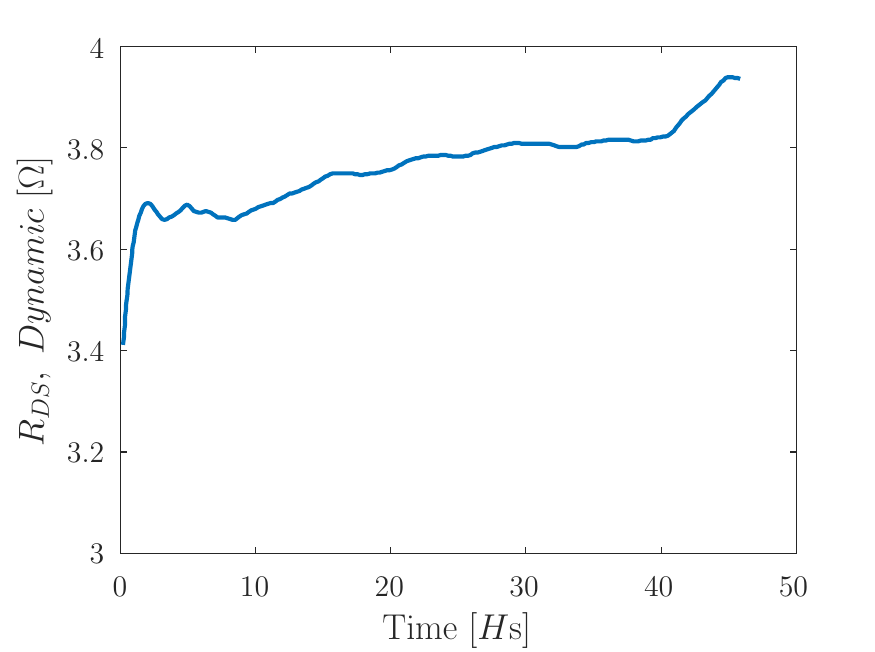}
\par\end{centering}
\caption{\label{fig11}$R_{DS(on)}$ versus time voltage 40v}

\end{figure}

\begin{figure}[h]
\begin{centering}
\includegraphics[scale=0.6]{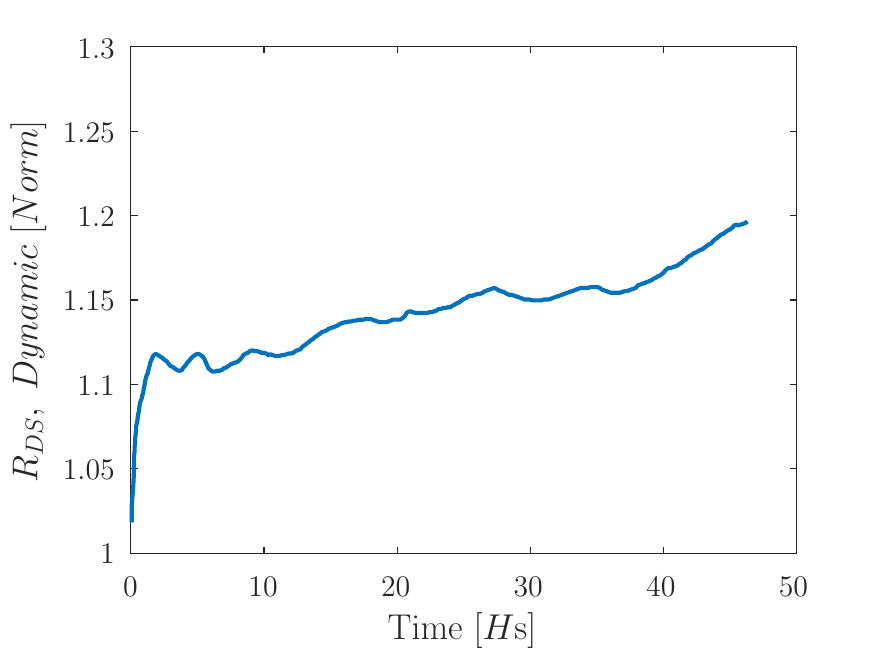}
\par\end{centering}
\caption{\label{fig:12}Normalized $R_{DS(on)}$ versus time voltage 40v}

\end{figure}

\begin{figure}[h]
\begin{centering}
\includegraphics[scale=0.6]{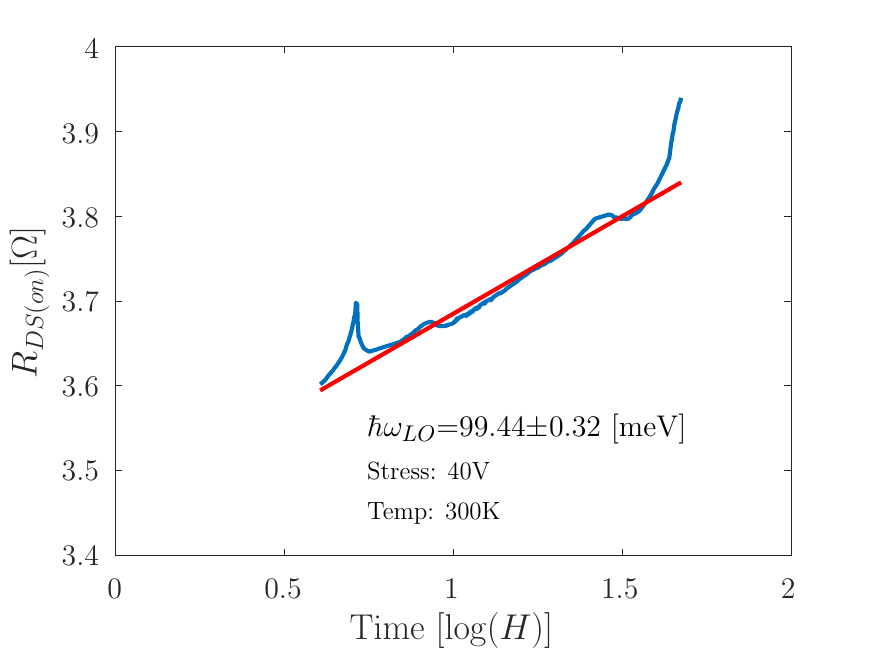}
\par\end{centering}
\caption{\label{fig:13}$R_{DS(on)}$ versus log time voltage 40v}

\end{figure}

As we can see from Figure \ref{fig11}, the graph of $R_{DS(on)}$
fits the log law of time, Figure \ref{fig:12}shows the Normalized
$R_{DS(on)}$ vs time and it also fits the log law of time. In Figure
\ref{fig:13} we had adjusted the time scale to log scale, as we can
see it fits a linear line which demonstrates that the $R_{DS(on)}$
follows the log law of time. In this experiment we did not successfully
validate the value of $\hbar\omega_{LO}$.

\subsubsection{Test with 70V stress, 400mA}

This test was performed with a constant drain current of 400 mA and
constant output voltage of 70v ,$V_{max}$ reached 85 v . Below in
Figure \ref{fig:14} we present the change of the $R_{DS(on)}$ during
this test which affected by the change of $V_{DS}$:
\begin{figure}[h]
\begin{centering}
\includegraphics[scale=0.6]{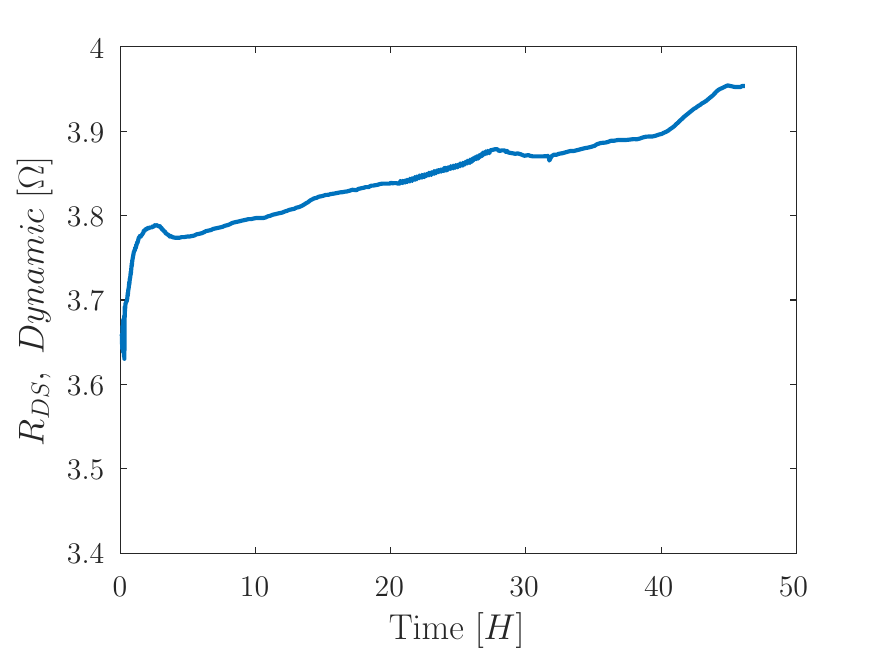}
\par\end{centering}
\caption{\label{fig:14}$R_{DS(on)}$ on versus time voltage 70v}

\end{figure}
\begin{figure}[h]
\begin{centering}
\includegraphics[scale=0.6]{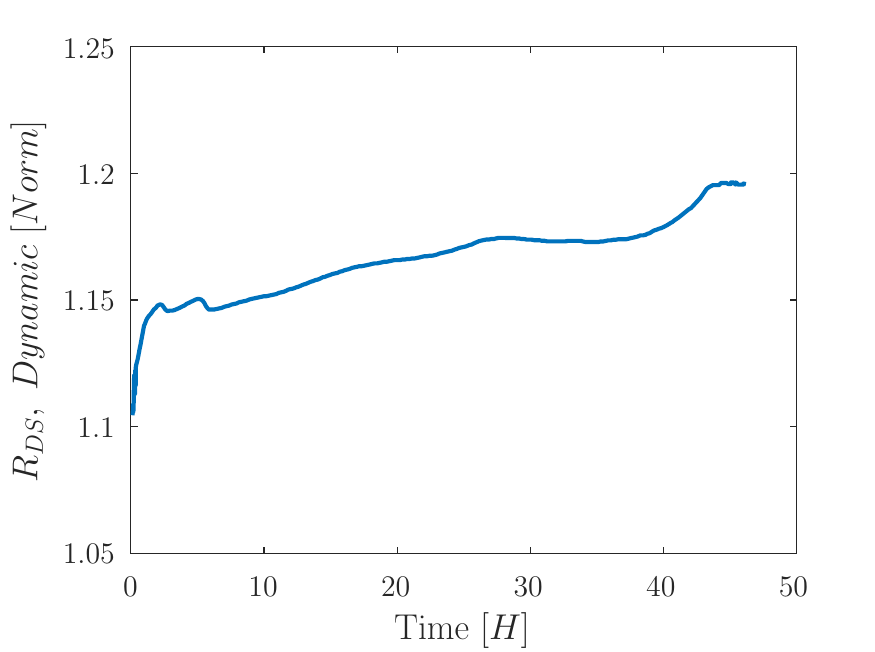}
\par\end{centering}
\caption{\label{fig:15}Normalized $R_{DS(on)}$ versus time voltage 70v}

\end{figure}

\begin{figure}[h]
\begin{centering}
\includegraphics[scale=0.6]{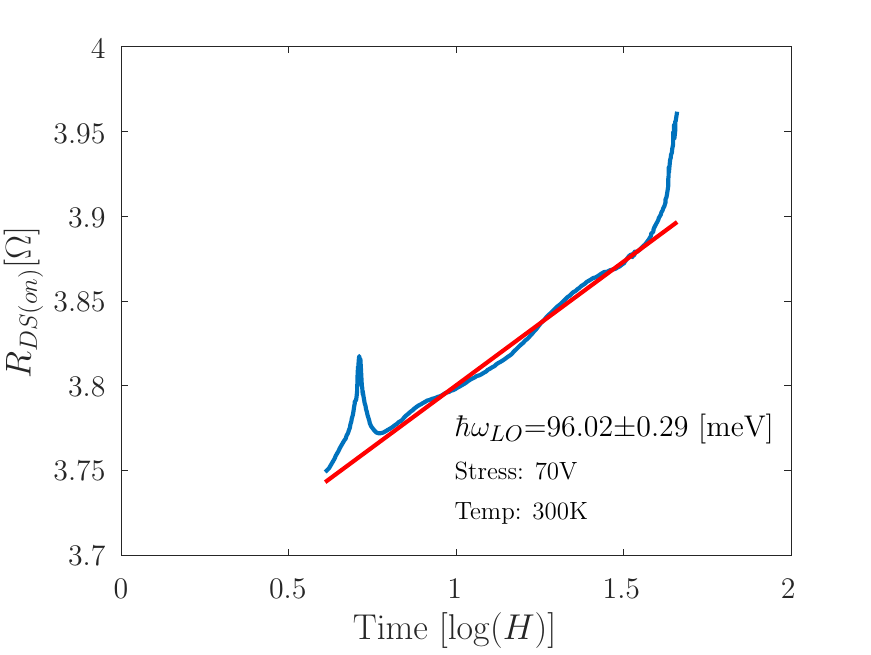}
\par\end{centering}
\caption{\label{fig:16}$R_{DS(on)}$ versus log time voltage 70v}

\end{figure}
As we can see from Figure \ref{fig:14}, the graph of $R_{DS(on)}$
fits the log law of time, Figure \ref{fig:15} shows the Normalized
$R_{DS(on)}$ vs time and it's also fit the log law of time. In Figure
\ref{fig:16} we had adjusted the time scale to log scale, as we can
see it fits a linear model which demonstrates that the $R_{DS(on)}$
follows a log law of time. This result is validated using the obtained
phonon energy in Karach's \cite{karch1997} work (see appendix).

\subsubsection{Test with 100V stress, 400mA}

This test was performed with a constant drain current of 400 mA and
constant output voltage of 100v ,$V_{max}$ reached 110 v ; below
in Figure \ref{fig:17} we present the change of the $R_{DS(on)}$
during this test which affected by the change of $V_{DS}$:
\begin{figure}[h]
\begin{centering}
\includegraphics[scale=0.6]{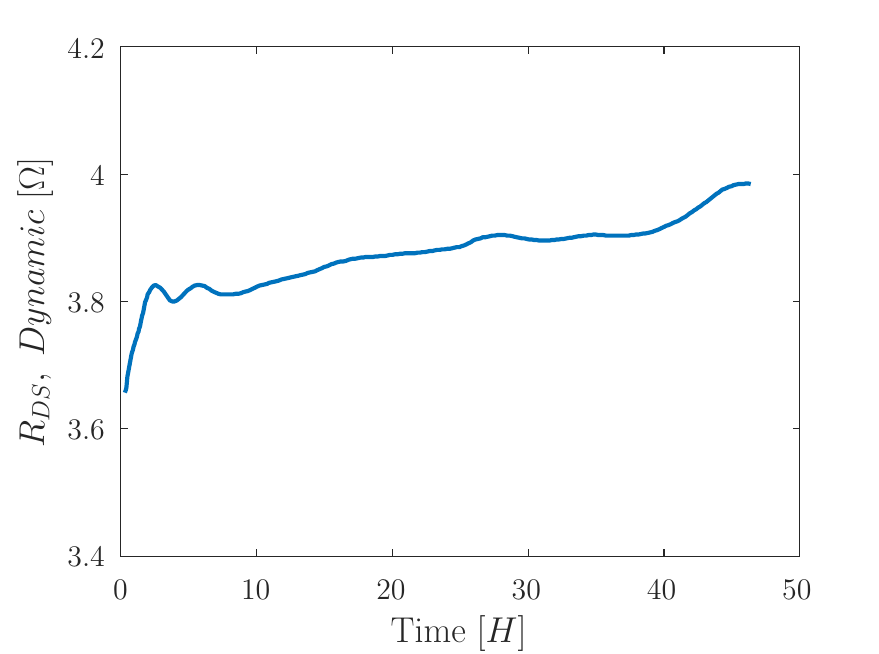}
\par\end{centering}
\caption{\label{fig:17}$R_{DS(on)}$ on versus time voltage 100v}

\end{figure}
\begin{figure}[h]
\begin{centering}
\includegraphics[scale=0.6]{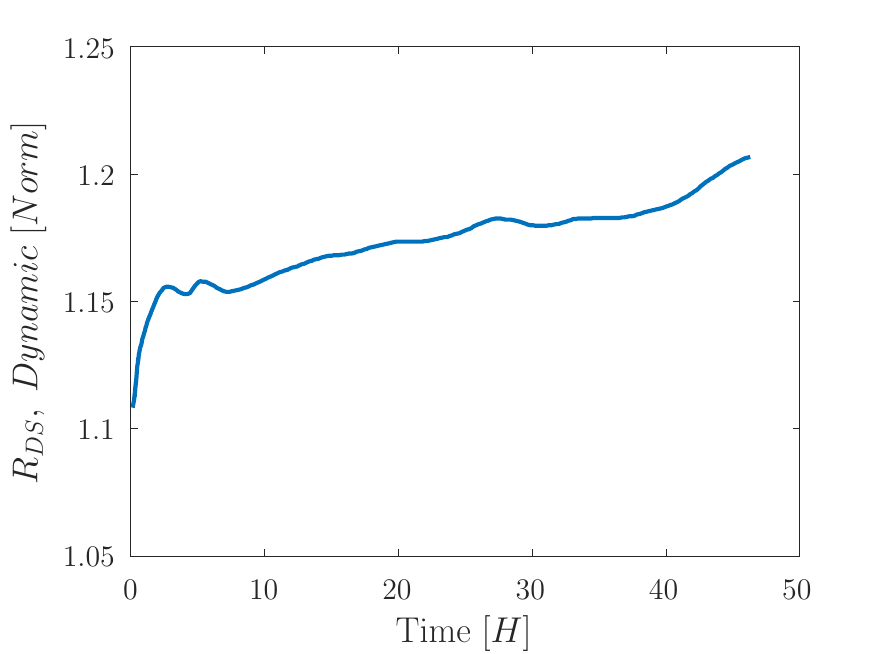}
\par\end{centering}
\caption{\label{fig:18}Normalized $R_{DS(on)}$ versus time voltage 100v}

\end{figure}
\begin{figure}[h]
\begin{centering}
\includegraphics[scale=0.6]{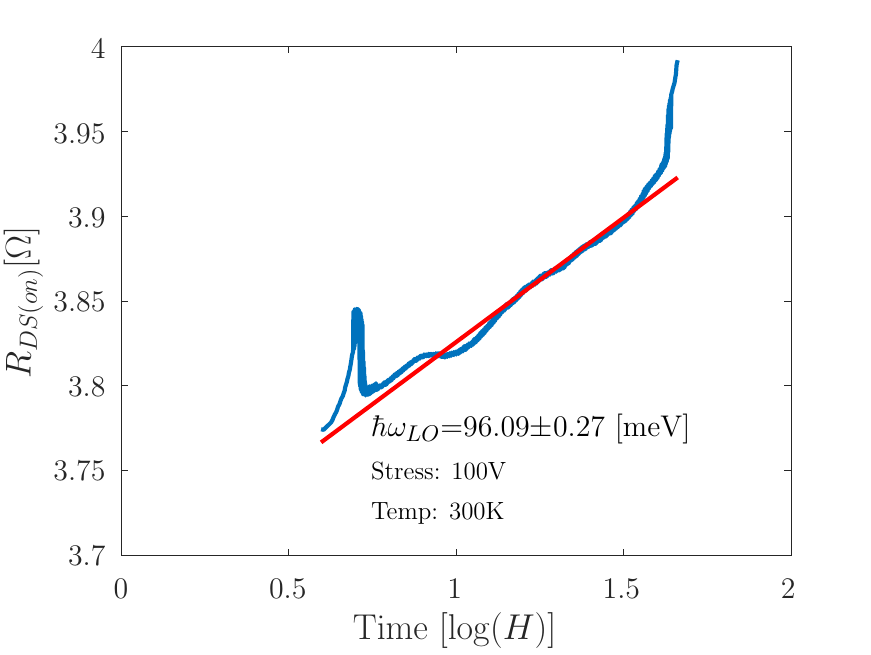}
\par\end{centering}
\caption{\label{fig:19}$R_{DS(on)}$ versus log time voltage 100v}

\end{figure}
As we can see from Figure \ref{fig:17}, the graph of $R_{DS(on)}$
fits the log law of time, Figure \ref{fig:18} shows the Normalized
$R_{DS(on)}$ vs time and it's also fit the log law of time. In Figure
\ref{fig:19} we had adjusted the time scale to log scale, as we can
see it fits a linear model which demonstrates that the $R_{DS(on)}$
follows a log law of time. This result is validated using the obtained
phonon energy in Karach's \cite{karch1997} work (see appendix).

\section{DISCUSSION}

In this research, a novel circuit and methodology were designed, developed,
and implemented to investigate the reliability of Gallium Nitride
(GaN) transistors under various stress conditions. By utilizing a
modified boost converter, the devices were tested across a range of
voltages, allowing for a detailed examination of degradation mechanisms.

Experimental measurements and calculations of the Drain-Source on-resistance
$R_{DS(on)}$ over time were conducted using the proposed circuit.
The results confirm that these changes follow a logarithmic law relative
to time, consistent with the findings reported in the EPC Phase 12
reliability report \cite{epc2020}.

The study utilized error propagation analysis to assist in qualifying
the test circuit by estimating the uncertainty in longitudinal optical
phonon scattering ($\hbar\omega_{LO}$). While initial tests at 40V
did not successfully validate the $\hbar\omega_{LO}$ value, subsequent
experiments at 70V and 100V yielded results that were validated against
existing theoretical data.

Moving forward, the goal is to further confirm the $R_{DS(on)}$ evolution
model developed by EPC using the methodology presented here. Additionally,
the research will aim to delve deeper into the underlying physics
of these findings to improve current physical models and our overall
understanding of GaN device reliability.

\bibliographystyle{ieeetr}
\bibliography{references}

@book{Meneghini2016PowerGaNDevices,
  title={Power {GaN} devices: Materials, applications and reliability},
  author={Meneghini, Matteo and Meneghesso, Gaudenzio and Zanoni, Enrico},
  year={2017},
  publisher={Springer}
}

@inproceedings{Gaska2014NovelAlInN,
  title = {Novel {AlInN / GaN} Integrated Circuits operating up to 500 $^\circ$C},
  author = {Gaska, R. and Gaevski, M. and Deng, J. and Jain, R.},
  booktitle = {Proc. Solid State Device Res. Conf. (ESSDERC), 44th Eur.},
  pages = {142--145},
  year = {2014}
}

@article{Herbecq20141900V,
  title = {1900V, $1.6m{\Omega}$ cm$^2$ {AlN/GaN-on-Si} power devices realized by local substrate removal},
  author = {Herbecq, N. and Roch-Jeune, I. and Rolland, N. and Visalli, D. and Derluyn, J. and Degroote, S. and Germain, M. and Medjdoub, F.},
  journal = {Appl. Phys. Express},
  volume = {7},
  pages = {034103},
  year = {2014}
}

@article{Chen2015RoomTemperatureMobility,
  title={Room-temperature mobility above 2200 $cm^2$/V{\textperiodcentered} s of two-dimensional electron gas in a sharp-interface {AlGaN/GaN} heterostructure},
  author={Chen, Jr-Tai and Persson, Ingemar and Nilsson, Daniel and Hsu, Chih-Wei and Palisaitis, Justinas and Forsberg, Urban and Persson, Per O{\AA} and Janz{\'e}n, Erik},
  journal={Applied Physics Letters},
  volume={106},
  number={25},
  year={2015},
  publisher={AIP Publishing}
}

@misc{GaNSystemsDatasheet,
  title = {GS66516T Top-side cooled 650 V E-mode {GaN} transistor Preliminary Datasheet},
  howpublished = {\url{http://www.gansystems.com/datasheets/GS66516T\%20DS\%20Rev\%20161007.pdf}}
}

@inproceedings{Moens2015ImpactOfBufferLeakage,
  title = {Impact of buffer leakage on intrinsic reliability of 650V {AlGaN / GaN HEMTs}},
  author = {Moens, P. and Banerjee, A. and Uren, M. J. and Meneghini, M. and Karboyan, S. and Chatterjee, I. and Vanmeerbeek, P. and Cäsar, M. and Liu, C. and Salih, A. and Zanoni, E. and Meneghesso, G. and Kuball, M. and Tack, M.},
  booktitle = {IEEE Electron Device Meet. IEDM 2015 Techical Dig.},
  pages = {903--906},
  year = {2015}
}

@article{Bisi2014KineticsOfBuffer,
  title = {Kinetics of Buffer-Related $R_{\text{ON}}$-Increase in {GaN-on-Silicon MISHEMTs}},
  author = {Bisi, D. and Meneghini, M. and Marino, F. A. and Marcon, D. and Stoffels, S. and Van Hove, M. and Decoutere, S. and Meneghesso, G. and Zanoni, E.},
  journal = {IEEE Electron Device Lett.},
  volume = {35},
  number = {10},
  pages = {1004--1006},
  year = {2014},
  doi = {10.1109/LED.2014.2346736}
}

@article{Meneghini2014OFFStateDegradation,
  title = {OFF-State Degradation of {AlGaN / GaN} Power {HEMTs}: Experimental Demonstration of time-dependent drain-source breakdown},
  author = {Meneghini, M. and Cibin, G. and Bertin, M. and Hurkx, A. G. M. and Ivo, P. and Šonský, J. and Croon, J. A. and Meneghesso, G. and Zanoni, E.},
  journal = {IEEE Trans. Electron Devices},
  volume = {61},
  number = {6},
  pages = {1987--1992},
  year = {2014},
  doi = {10.1109/TED.2014.2319011}
}

@article{Rossetto2017FieldRelatedFailure,
  title = {Field-Related Failure of {GaN-on-Si HEMTs}: Dependence on Device Geometry and Passivation},
  author = {Rossetto, I. and Meneghini, M. and Pandey, S. and Gajda, M. and Hurkx, G. A. M. and Croon, J. A. and Šonský, J. and Meneghesso, G. and Zanoni, E.},
  journal = {IEEE Trans. Electron Devices},
  volume = {64},
  number = {1},
  pages = {73--77},
  year = {2017},
  doi = {10.1109/TED.2016.2625290}
}

@inproceedings{Wu2015TimeDependentDielectric,
  title = {Time dependent dielectric breakdown ({TDDB}) evaluation of {PE-ALD SiN} gate dielectrics on {AlGaN/GaN} recessed gate {D}-mode {MISHEMTs} and E-mode {MIS-FETs}},
  author = {Wu, T. and Marcon, D. and De Jaeger, B. and Van Hove, M. and Bakeroot, B. and Stoffels, S. and Groeseneken, G. and Decoutere, S. and Roelofs, R.},
  booktitle = {Reliab. Phys. Symp. (IRPS), 2015 IEEE Int.},
  pages = {6C.4.1},
  year = {2015}
}

@article{Uemoto2007GateInjectionTransistor,
  title = {Gate injection transistor ({GIT}) - {A} normally-off {AlGaN/GaN} power transistor using conductivity modulation},
  author = {Uemoto, Y. and Hikita, M. and Ueno, H. and Matsuo, H. and Ishida, H. and Yanagihara, M. and Ueda, T. and Tanaka, T. and Ueda, D.},
  journal = {IEEE Trans. Electron Devices},
  volume = {54},
  number = {12},
  pages = {3393--3399},
  year = {2007},
  doi = {10.1109/TED.2007.905499}
}

@article{tapajna2015investigation,
  title = {Investigation of gate-diode degradation in normally-off p-{GaN / AlGaN / GaN} high-electron-mobility transistors},
  author = {{\v{T}}apajna, M and Hilt, O and Bahat-Treidel, E and W{\"u}rfl, J and Kuzm{\'\i}k, J},
  journal = {Appl. Phys. Lett.},
  volume = {107},
  pages = {193506},
  year = {2015},
  doi = {10.1063/1.4935216}
}

@article{Wu2013ComprehensiveInvestigation,
  title = {Comprehensive Investigation of On-State Stress on D- Mode {AlGaN / GaN MIS-HEMTs}},
  author = {Wu, T. and Marcon, D. and Zahid, M. B. and Van Hove, M. and Decoutere, S. and Groeseneken, G.},
  journal = {IEEE Int. Reliab. Phys. Symp.},
  pages = {1--7},
  year = {2013},
  doi = {10.1109/IRPS.2013.6531920}
}

@inproceedings{Joh2014CurrentCollapse,
  title = {Current collapse in {GaN} heterojunction field effect transistors for highvoltage switching applications},
  author = {Joh, J. and Tipirneni, N. and Pendharkar, S. and Krishnan, S.},
  booktitle = {IEEE Int. Reliab. Phys. Symp. Proc.},
  pages = {4--7},
  year = {2014}
}

@inproceedings{Bahl2016ProductLevelReliability,
  title = {Product-level Reliability of {GaN} Devices},
  author = {Bahl, S. R. and Ruiz, D. and Lee, D. S.},
  booktitle = {IEEE Int. Reliab. Phys. Symp.},
  year = {2016}
}

@article{golan2018improved,
  title={An improved reliability model for {Si} and {GaN} power {FET}},
  author={Golan, Gady and Azoulay, Moshe and Avraham, Tsuriel and Kremenetsky, Ilan and Bernstein, Joseph B},
  journal={Microelectronics Reliability},
  volume={81},
  pages={77--89},
  year={2018},
  publisher={Elsevier}
}

@article{Gao2011RoleOfOxygen,
  title = {Role of oxygen in the off-state degradation of {AlGaN/GaN} high electron mobility transistors},
  author = {Gao, F. and Lu, B. and Li, L. and et al.},
  journal = {Appl. Phys. Lett.},
  year = {2011},
  volume = {99},
  number = {22},
  pages = {223506},
  doi = {10.1063/1.3665065}
}

@article{Gao2014ImpactOfWaterAssisted,
  title = {Impact of water-assisted electrochemical reactions on the off-state degradation of {AlGaN/GaN HEMTs}},
  author = {Gao, F. and Tan, S.C. and del Alamo, J.A. and et al.},
  journal = {IEEE Trans. Electron Devices},
  year = {2014},
  volume = {61},
  number = {2},
  pages = {437--444},
  doi = {10.1109/TED.2013.2293114}
}

@article{Ambacher2002PyroelectricProperties,
  title = {Pyroelectric properties of {Al(In)GaN/GaN} hetero- and quantum well structures},
  author = {Ambacher, O. and Majewski, J. and Miskys, C. and et al.},
  journal = {J. Phys. Condens. Matter},
  year = {2002},
  volume = {14},
  number = {13},
  pages = {3399--3434},
  doi = {10.1088/0953-8984/14/13/302}
}

@article{Joh2008CriticalVoltage,
  title = {Critical voltage for electrical degradation of {GaN} high-electron mobility transistors},
  author = {Joh, J. and del Alamo, J.A.},
  journal = {IEEE Electron Device Lett.},
  year = {2008},
  volume = {29},
  number = {4},
  pages = {287--289},
  doi = {10.1109/LED.2008.917815}
}

@inproceedings{Marcon2010ComprehensiveReliability,
  title = {A comprehensive reliability investigation of the voltage-, temperature- and device geometry-dependence of the gate degradation on state-of-the-art {GaN}-on-{Si} {HEMTs}},
  author = {Marcon, D. and Kauerauf, T. and Medjdoub, F. and et al.},
  booktitle = {2010 Int. Electron Devices Meeting},
  pages = {20.3.1--20.3.4},
  year = {2010},
  doi = {10.1109/IEDM.2010.5703398}
}

@article{Suehle2002UltrathinGateOxide,
  title = {Ultrathin gate oxide reliability: physical models, statistics, and characterization},
  author = {Suehle, J.S.},
  journal = {IEEE Trans. Electron Devices},
  year = {2002},
  volume = {49},
  number = {6},
  pages = {958--971},
  doi = {10.1109/TED.2002.1003712}
}

@article{inberg2001material,
  title={Material and electrical properties of electroless {Ag-W} thin film},
  author={Inberg, A and Shacham-Diamand, Y and Rabinovich, E and Golan, G and Croitoru, N},
  journal={Journal of electronic materials},
  volume={30},
  number={4},
  pages={355--359},
  year={2001},
  publisher={Springer}
}

@article{Rossetto2016TimeDependentFailure,
  title = {Time-dependent failure of {GaN-on-Si} power {HEMTs} with {p-GaN} gate},
  author = {Rossetto, I. and Meneghini, M. and Hilt, O. and et al.},
  journal = {IEEE Trans. Electron Devices},
  year = {2016},
  volume = {63},
  number = {6},
  pages = {2334--2339},
  doi = {10.1109/TED.2016.2553721}
}

@article{Rossetto2016StudyOfTheStability,
  title = {Study of the stability of e-mode {GaN HEMTs} with {p-GaN} gate based on combined {DC} and optical analysis},
  author = {Rossetto, I. and Meneghini, M. and Rizzato, V. and et al.},
  journal = {Microelectron. Reliab.},
  year = {2016},
  volume = {64},
  pages = {547--551},
  doi = {10.1016/j.microrel.2016.07.127}
}

@article{Meneghini2015ExtensiveInvestigation,
  title = {Extensive investigation of time-dependent breakdown of {GaN-HEMTs} submitted to off-state stress},
  author = {Meneghini, M. and Rossetto, I. and Hurkx, F. and et al.},
  journal = {IEEE Trans. Electron Devices},
  year = {2015},
  volume = {62},
  number = {8},
  pages = {2549--2554},
  doi = {10.1109/TED.2015.2446032}
}

@article{Degraeve1998NewInsights,
  title = {New insights in the relation between electron trap generation and the statistical properties of oxide breakdown},
  author = {Degraeve, R. and Groeseneken, G. and Bellens, R. and et al.},
  journal = {IEEE Trans. Electron Devices},
  year = {1998},
  volume = {45},
  number = {4},
  pages = {904--911},
  doi = {10.1109/16.662800}
}

@article{Stathis1999PercolationModels,
  title = {Percolation models for gate oxide breakdown},
  author = {Stathis, J.H.},
  journal = {J. Appl. Phys.},
  year = {1999},
  volume = {86},
  number = {10},
  pages = {5757},
  doi = {10.1063/1.371590}
}

@article{golan2000novel,
  title={Novel sputtering method for {Pd--Al2O3 UV} transparent conductive coatings},
  author={Golan, G and Axelevitch, A},
  journal={Microelectronics Journal},
  volume={31},
  number={6},
  pages={469--473},
  year={2000},
  publisher={Elsevier}
}

@article{Meneghini2016NormallyOffGaN,
  title = {Normally-off {GaN-HEMTs} with p-type gate: off-state degradation, forward gate stress and {ESD} failure},
  author = {Meneghini, M. and Hilt, O. and Fleury, C. and et al.},
  journal = {Microelectron. Reliab.},
  year = {2016},
  volume = {58},
  pages = {177--184},
  doi = {10.1016/j.microrel.2015.11.023}
}

@article{Rossetto2015DemonstrationOfField,
  title={Demonstration of field and power dependent {ESD} failure in {AlGaN/GaN RF HEMTs}},
  author={Rossetto, Isabella and Meneghini, Matteo and Barbato, Marco and Rampazzo, Fabiana and Marcon, Denis and Meneghesso, Gaudenzio and Zanoni, Enrico},
  journal={IEEE Transactions on Electron Devices},
  volume={62},
  number={9},
  pages={2830--2836},
  year={2015},
  publisher={IEEE}
}

@article{Kuzmik2004ElectricalOverstress,
  title={Electrical overstress in {AlGaN/GaN HEMTs}: Study of degradation processes},
  author={Kuzm{\'i}k, J and Pog{\'a}ny, Dionyz and Gornik, Erich and Javorka, Peter and Kordo{\v{s}}, P},
  journal={Solid-State Electronics},
  volume={48},
  number={2},
  pages={271--276},
  year={2004},
  publisher={Elsevier}
}

@article{Chen2012HBMEsdRobustness,
  title = {{HBM ESD} robustness of {GaN-on-Si} {S}chottky diodes},
  author = {Chen, S.H and Griffoni, A. and Srivastava, P. and et al.},
  journal = {IEEE Trans. Device Mater. Reliab.},
  year = {2012},
  volume = {12},
  number = {4},
  pages = {589--598},
  doi = {10.1109/TDMR.2012.2217746}
}

@inproceedings{Shankar2016UniqueEsdBehavior,
  title = {Unique {ESD} behavior and failure modes of {AlGaN/GaN HEMTs}},
  author = {Shankar, B. and Shrivastava, M.},
  booktitle = {2016 IEEE Int. Reliability Physics Symp. (IRPS)},
  pages = {EL-7-1--EL-7-5},
  year = {2016},
  doi = {10.1109/IRPS.2016.7574608}
}

@article{Khan1994CurrentVoltageCharacteristic,
  title = {Current/voltage characteristic collapse in {AlGaN/GaN} heterostructure insulated gate field effect transistors at high drain bias},
  author = {Khan, M. A. and Shur, M. S. and Chen, Q. C. and Kuznia, J. N.},
  journal = {Electron. Lett.},
  volume = {30},
  number = {25},
  pages = {2175--2176},
  month = dec,
  year = {1994}
}

@article{Vetury2001ImpactOfSurfaceStates,
  title = {The impact of surface states on the {DC} and {RF} characteristics of {AlGaN/GaN HFETs}},
  author = {Vetury, R. and Zhang, N. Q. and Keller, S. and Mishra, U. K.},
  journal = {IEEE Trans. Electron Devices},
  volume = {48},
  number = {3},
  pages = {560--566},
  month = mar,
  year = {2001}
}

@article{Hu2001Si3N4AlGaN,
  title = {Si$_3${N}$_4$/{AlGaN}/{GaN}-metal-insulator-semiconductor heterostructure field-effect transistors},
  author = {Hu, X. and Koudymov, A. and Simin, G. and Yang, J. and Khan, M. A. and Tarakji, A. and Shur, M. S. and Gaska, R.},
  journal = {Appl. Phys. Lett.},
  volume = {79},
  number = {17},
  pages = {2832--2834},
  month = oct,
  year = {2001}
}

@article{Leirer2001RFCharacterization,
  title = {{RF} characterization and transit behavior of {AlGaN/GaN} power {HFETs}},
  author = {Leirer, H. and Vescan, A. and Dietrich, R. and Wieszt, A. and Sledzik, H. H.},
  journal = {IEICE Trans. Electron.},
  volume = {E84-C},
  number = {10},
  pages = {1442--1447},
  year = {2001}
}

@article{Kim2003EffectsOfSiNPassivation,
  title = {Effects of {SiN} passivation and high-electric field on {AlGaN-GaN HFET} degradation},
  author = {Kim, H. and Thompson, R. M. and Tilak, V. and Prunty, T. R. and Shealy, J. R. and Eastman, L. F.},
  journal = {IEEE Electron Device Lett.},
  volume = {24},
  number = {7},
  pages = {421--423},
  month = jul,
  year = {2003}
}

@article{Ibbetson2000PolarizationEffects,
  title = {Polarization effects, surface states, and the source of electrons in {AlGaN/GaN} heterostructure field effect transistors},
  author = {Ibbetson, J. P. and Fini, P. T. and Ness, K. D. and DenBaars, S. P. and Speck, J. S. and Mishra, U. K.},
  journal = {Appl. Phys. Lett.},
  volume = {77},
  number = {2},
  pages = {250--252},
  month = jul,
  year = {2000}
}

@article{valizadeh2005effects,
  title={Effects of {RF} and {DC} stress on {AlGaN/GaN MODFETs}: A low-frequency noise-based investigation},
  author={Valizadeh, Pouya and Pavlidis, Dimitris},
  journal={IEEE Transactions on Device and Materials Reliability},
  volume={5},
  number={3},
  pages={555--563},
  year={2005},
  publisher={IEEE}
}

@article{axelevitch2014application,
  title={Application of gold nano-particles for silicon solar cells efficiency increase},
  author={Axelevitch, A and Gorenstein, B and Golan, G},
  journal={Applied surface science},
  volume={315},
  pages={523--526},
  year={2014},
  publisher={Elsevier}
}

@article{orr2025gallium,
  title={Gallium {N}itride {H}igh-{E}lectron-{M}obility {T}ransistor-{B}ased {H}igh-{E}nergy {P}article-{D}etection {P}reamplifier},
  author={Orr, Gilad and Azoulay, Moshe and Golan, Gady and Burger, Arnold},
  journal={Metrology},
  volume={5},
  number={2},
  pages={21},
  year={2025},
  publisher={MDPI}
}

@article{orr2012safe,
  title={Safe and consistent method of spot-welding platinum thermocouple wires and foils for high temperature measurements},
  author={Orr, Gilad and Roth, Michael},
  journal={Review of scientific instruments},
  volume={83},
  number={8},
  year={2012},
  publisher={AIP Publishing}
}

@techreport{epc2020,
  author = {{EPC Corporation}},
  title = {GaN Reliability Report Phase 12},
  year = {2021},
  url = {https://epc-co.com/epc/portals/0/epc/documents/product-training/Reliability%20Report%20Phase%2012.pdf}
}

@article{fang2019electron,
  title={Electron Transport Properties of Al x Ga 1- x N/Ga N Transistors Based on First-Principles Calculations and Boltzmann-Equation Monte Carlo Simulations},
  author={Fang, Jingtian and Fischetti, Massimo V and Schrimpf, Ronald D and Reed, Robert A and Bellotti, Enrico and Pantelides, Sokrates T},
  journal={Physical Review Applied},
  volume={11},
  number={4},
  pages={044045},
  year={2019},
  publisher={APS}
}

@article{davydov1998,
  title = {Phonon dispersion and Raman scattering in hexagonal GaN and AIN},
  author = {Davydov, V. Yu. and Kitaev, Yu. E. and Goncharuk, I. N. and Smirnov, A. N. and Graul, J. and Semchinova, O. and Uffmann, D. and Smirnov, M. B. and Mirgorodsky, A. P. and Evarestov, R. A.},
  journal = {Physical Review B},
  volume = {58},
  pages = {12899},
  year = {1998},
  doi = {10.1103/PhysRevB.58.12899}
}

@article{bungaro2000,
  title = {Ab initio phonon dispersions of wurtzite AIN, GaN, and InN},
  author = {Bungaro, Claudia and Rapcewicz, Krzysztof and Bernholc, J.},
  journal = {Physical Review B},
  volume = {61},
  pages = {6720},
  year = {2000},
  doi = {10.1103/PhysRevB.61.6720}
}

@article{karch1997,
  title = {Ab initio study of lattice dynamics of GaN, AlN, and BN},
  author = {Karch, K. and Wagner, J. M. and Bechstedt, F.},
  journal = {Physical Review B},
  volume = {57},
  pages = {7043},
  year = {1998},
  doi = {10.1103/PhysRevB.57.7043}
}

@article{perlin1992,
  title = {Raman scattering and x-ray-absorption spectroscopy in gallium nitride under high pressure},
  author = {Perlin, P. and Jauberthie-Carillon, C. and Itie, J. P. and Miguel, A. San and Grzegory, I. and Polian, A.},
  journal = {Physical Review B},
  volume = {45},
  number = {1},
  pages = {83--89},
  year = {1992},
  publisher = {American Physical Society},
  doi = {10.1103/PhysRevB.45.83}
}

\appendix{\noindent \textsc{The LO Phonon Energy in Wurtzite GaN}}

\setcounter{equation}{0} 
\setcounter{table}{0} 
\setcounter{figure}{0}

\renewcommand{\theequation}{A.\arabic{equation}}
\renewcommand{\thetable}{A.\Roman{table}}
\renewcommand{\thefigure}{A.\arabic{figure}}

For validating the results of this work, determining the energy of
the longitudinal optical (LO) phonon and its uncertainty in GaN is
of sought importance. While the EPC Report \cite{epc2020}sites the
accepted value of approximately 92 meV, as we shall see, high resolution
Raman spectroscopy and first principles calculations provide a range
of values between 91.4 and 93.9 meV.

In Raman spectroscopy, phonon energies are measured in inverse centimeters
($cm^{-1})$ hence the findings need to be converted to electronvolts
(eV). As $E=\frac{hc}{\lambda}$ the relationship is:
\begin{equation}
E(eV)=\frac{h\cdot c\cdot\tilde{\nu}}{e}
\end{equation}
where $h$ is Planck's constant, $c$ the speed of light, $\tilde{\nu}$
the wavenumber in $cm^{-1}$ and $e$ the elementary charge. The conversion
in eV is
\begin{equation}
E(eV)=\frac{\tilde{\nu}}{8065.54}\,[\text{eV}]
\end{equation}

the 92 meV figure commonly discussed in GaN literature corresponds
to a wavenumber of approximately $742-744\,cm^{-1}$ observed via
Raman Scattering. Raman Scattering is the primary tool for measuring
phonon frequencies at the Brillouin zone center ($\Gamma$ point)
as other traditional methods fail for large GaN crystals are challenging
to grow.

Davydov et al. \cite{davydov1998} reported the measured $E_{1}(LO)$
mode at $744.0\pm0.2\,cm^{-1}$ which corresponds to $92.24\pm0.025\,meV$.

Perlin et al. \cite{perlin1992} which is widely cited, reported the
mode at $742\,cm^{-1}$ without citing the uncertainty corresponding
to $92.0\,meV$.

The energy of the longitudinal optical phonon was calculated theoretically
with values depending on the mathematical approximations used.

Bungaro et al. \cite{bungaro2000} calculated the $E_{1}(LO)$ mode
at $737\pm1\,cm^{-1}$ using DFT methodsestimating the uncertainty
based on their convergence error. This corresponds to an energy of
$91.38\pm0.12\,meV$. 

Karach et al. \cite{karch1997} Calculated the $E_{1}(LO)$ mode at
$757\,cm^{-1}$ he states that there is an inherent 3\% "error"
caused by approximations in the DFT itself which corresponds to $93.86\pm2.8\,meV$.
The importance of this result is that it is validated by Davydov's
experiment and it \emph{validates our results as well}.

The following table compares the various published results of GaN
$E_{1}(LO)$ phonon energy values with their uncertainties if available.

\begin{table*}[h!] \centering \caption{Comparison of $E_1(LO)$ Phonon Energies in Wurtzite GaN} \label{tab:gan_lo_phonons} \begin{tabular}{l l c c c} \hline \hline \textbf{Source} & \textbf{Type} & \textbf{Freq. ($\text{cm}^{-1}$)} & \textbf{Energy (meV)} & \textbf{Uncertainty (meV)} \\ [0.5ex]  \hline Karch et al. (1998) \cite{karch1997} & Theoretical (DFT) & 757 & 93.86 & $\approx \pm 2.8^a$ \\  Davydov et al. (1998) \cite{davydov1998} & Experimental (Raman) & 744.0 & 92.24 & $\pm 0.025^b$ \\ Perlin et al. (1992) \cite{perlin1992} & Experimental (Raman) & 742 & 92.00 & --  \\ Bungaro et al. (2000) \cite{bungaro2000} & Theoretical (DFPT) & 737 & 91.38 & $\pm 0.12^c$ \\ [1ex]  \hline \hline \end{tabular} \begin{flushleft} \footnotesize{$^a$ Reflects the standard systematic uncertainty of $\approx 3\%$ for LDA-based DFT models.}\\ \footnotesize{$^b$ Reported instrumental precision of $\pm 0.2 \text{ cm}^{-1}$.}\\ \footnotesize{$^c$ Numerical convergence limit of $\pm 1 \text{ cm}^{-1}$ reported by the authors.} \end{flushleft} \end{table*}
\end{document}